\documentclass{aa}
\usepackage{graphicx}
\usepackage{txfonts}
\usepackage[]{natbib}
\def \chisq {$\chi^2$}
\begin{document}
   \title{GRB~070311: a direct link between the prompt emission and the afterglow}

%   \subtitle{ }

   \author{C.~Guidorzi\inst{1,2}, S.~D. Vergani\inst{3,4},
S.~Sazonov\inst{5,6}, S.~Covino\inst{2}, D.~Malesani\inst{7}, S.~Molkov\inst{6}, E.~Palazzi\inst{8},
P.~Romano\inst{1,2}, S.~Campana\inst{2}, G.~Chincarini\inst{1,2},
D.~Fugazza\inst{1,2}, A.~Moretti\inst{2}, G.~Tagliaferri\inst{2},
A.~Llorente\inst{9}, J.~Gorosabel\inst{10}, L.A.~Antonelli\inst{11,12}, M.~Capalbi\inst{12},
G.~Cusumano\inst{13},
P.~D'Avanzo\inst{2,14}, V.~Mangano\inst{13}, N.~Masetti\inst{8}, E.~Meurs\inst{3,4}, T.~Mineo\inst{13},
E.~Molinari\inst{2}, D.C.~Morris\inst{15}, L.~Nicastro\inst{8}, K.L.~Page\inst{16}, M.~Perri\inst{12},
B.~Sbarufatti\inst{13}, G.~Stratta\inst{12}, R.~Sunyaev\inst{5,6}, E.~Troja\inst{13,16}, F.M.~Zerbi\inst{2}
}

   \offprints{C.\ Guidorzi, cristiano.guidorzi@brera.inaf.it}
   \institute{Universit\`a degli studi di Milano-Bicocca,
                 Dipartimento di Fisica, piazza delle Scienze 3, 
                 I-20126 Milano, Italy
    \and INAF -- Osservatorio Astronomico di Brera, 
                 via E. Bianchi 46, I-23807 Merate (LC), Italy
    \and Dunsink Observatory - DIAS, 31 Fitzwilliam Street, Dublin 2, Ireland
    \and School of Physical Sciences and NCPST, Dublin City University, Dublin 9, Ireland
    \and Max-Planck-Institut f\"ur Astrophysik, Karl-Schwarzschild-Str. 1,
85740 Garching bei M\"unchen, Germany
    \and Space Research Institute, Russian Academy of Sciences,
Profsoyuznaya 84/32, I-117997 Moscow, Russia
    \and Dark Cosmology Centre, Niels Bohr Institute, University of Copenhagen, Juliane Maries
vej 30, DK-2100 K\o behavn \O , Denmark
    \and INAF -- Istituto di Astrofisica Spaziale e Fisica Cosmica, Sezione di Bologna, via P. Gobetti 101,
I-40129 Bologna, Italy
    \and XMM-Newton Science Operations Centre, European Space Agency, Villafranca del Castillo,
PO Box 50727, I-28080 Madrid, Spain
    \and Instituto de Astrof\'isica de Andaluc\'ia (IAA-CSIC), Apartado de Correos, 3.004, I-18080
Granada, Spain
    \and INAF -- Osservatorio Astronomico di Roma, via Frascati 33, I-00040, Monte Porzio Catone, Italy
    \and ASI Science Data Center, via G. Galilei, I-00044 Frascati (Roma), Italy\thanks{INAF personnel resident at ASDC}
    \and INAF--Istituto di Astrofisica Spaziale e Fisica Cosmica, Sezione di Palermo,
via U. La Malfa 153, I-90146 Palermo, Italy
    \and Universit\`a dell'Insubria, Dipartimento di Fisica e Matematica, via Valleggio 11,
I-22100 Como, Italy
    \and Department of Astronomy and Astrophysics, Pennsylvania State University, University Park, PA
    \and Department of Physics and Astronomy, University of Leicester, Leicester, LE1 7RH, UK
             }

  % \date{Received 1 March 2007/Accepted}
   \date{\today}
%
% \abstract{}{}{}{}{}
% 5 {} token are mandatory
\abstract
{ The prompt emission of gamma-ray bursts is mostly thought to be produced in internal shocks
of relativistic shells emitted by the progenitor at different times, whereas the late multi-band afterglow
is interpreted as the synchrotron emission of electrons swept up by the fireball expanding through
the surrounding interstellar medium. The short timescale variability observed in flares superimposed
on the X-ray/optical afterglow of several bursts recently made possible by Swift,
has been interpreted as evidence for prolonged activity of the inner engine through internal shocks.
Yet, it is not clear whether this applies to all the observed bursts and, in particular, whether
the bursts exhibiting single $\gamma$-ray pulses with no short timescale variability at late
times could also be entirely interpreted as external shocks.}
%%%%%%%%%%
{We present prompt $\gamma$-ray, early NIR/optical, late optical and X-ray observations of the
peculiar GRB~070311 discovered by {\rm INTEGRAL}, in order to gain clues on the mechanisms
responsible for the prompt $\gamma$-ray pulse as well as for the early and late multi-band afterglow
of GRB~070311.}
%%%%%%%%%%
{We fitted with empirical functions the gamma-ray and optical light curves and scaled
the result to the late time X-rays.}
%%%%%%%%%%
{The $H$-band light curve taken by REM shows two pulses peaking 80 and
140~s after the peak of the $\gamma$-ray burst and possibly accompanied by a faint $\gamma$-ray tail.
Remarkably, the late optical and X-ray afterglow underwent
a major rebrightening between $3\times10^4$ and $2\times10^5$~s after
the burst with an X-ray fluence comparable with that of the prompt emission extrapolated in the same band.
Notably, the time profile of the late rebrightening can be described as the combination of a
time-rescaled version of the prompt $\gamma$-ray pulse and an underlying power law.}
%%%%%%%%%%
{This result supports a common origin for both prompt and late X-ray/optical afterglow
rebrightening of GRB~070311 within the external shock scenario.
The main fireball would be responsible for the prompt emission, while a second
shell would produce the rebrightening when impacting the leading blastwave in a
refreshed shock.}

% heading (mandatory)
{}
\authorrunning {C.\ Guidorzi  et al.}
\titlerunning {Prompt and afterglow study of GRB~070311}
\maketitle

\section{Introduction}\label{sec:intro}
The bewildering variety of the long $\gamma$-ray prompt emission profiles
of gamma-ray bursts (GRBs) has been puzzling astronomers since their discovery
(e.g. see Fishman \& Meegan 1995 for a review).\nocite{Fishman95}
Among the most important open issues, two are still nowadays debated:
first, how long does the central engine remain active? Does the emission consist of a single
episode or a temporal sequence of events with interspersed quiescent periods?
Secondly, after the discovery of the delayed multi-band long-lasting emission called
``afterglow'', made possible for the first time by {\rm BeppoSAX} ten years ago
\citep{Costa97}, a general consensus on the emission mechanisms of both the prompt
and the afterglow emission is still missing, even though for the latter synchrotron
radiation by a population of shocked electrons proved to be successful
in accounting for a number of spectral and temporal evolution properties (e.g., 
M\'esz\'aros 2006).\nocite{Meszaros06}
Gaining clues on these issues may help to shed light on the nature of the progenitors as
well as on the circumburst environment.

The advent of {\rm Swift} \citep{Gehrels04} has been allowing the study of
multi-band afterglows as early as $\sim10^2$~s after the prompt event, exploring a previously
unknown time domain.
Both optical and X-ray light curves have shown unexpected behaviours that still lack a solid
agreed theoretical explanation.
Among the impressive discoveries by {\rm Swift} concerning the long GRBs, here we mention
the canonical behaviour of steep-shallow-normal decay characterising most of the early X-ray
afterglows as well as the occurrence of X-ray flares in $\sim50$\% of them (see Zhang 2007,
for a review up to date).\nocite{Zhang07_rev}

A number of different interpretations of the shallow decay phase experienced by the X-ray afterglows
of a sizable fraction of {\rm Swift} bursts \citep{Tagliaferri05,OBrien06,Nousek06,Zhang06} have been proposed.
They can be broadly classified according to which mechanism is invoked: internal (IS;
e.g., Rees \& M\'esz\'aros 1994; Kobayashi et al. 1997)\nocite{Rees94,Kobayashi97}
or external shocks (ES; e.g., Rees \& M\'esz\'aros 1992; Shaviv \& Dar 1995; Fenimore et al. 1996;
Dermer \& Mitman 1999).\nocite{Rees92,Shaviv95,Fenimore96,Dermer99}
For instance, within the IS model the shallow decay phase could be the result of low
velocity contrast distribution wind of slow shells emitted soon after the fast ones \citep{Granot06}.
According to alternative ES interpretations, it could be the result of delayed energy injection
into the fireball, either in the form of freshly ejected material from late engine activity
(``refreshed shock scenario''; Rees \& M\'esz\'aros 1998),\nocite{Rees98}
or by the collision of a wind of low velocity contrast shells catching up with the fireball
\citep{Zhang01}.
More generally, the shallow decay could be the result of a ``late prompt'' emission,
i.e. the same mechanism responsible for the $\gamma$-ray pulses
of the prompt would be at work through later shells with decreasing bulk Lorentz
factors \citep{Ghisellini07}.
Overall, a consensus is still missing,
also because of the large variety of behaviours observed:
in some cases, there is no spectral
evolution across the break in the light curve marking the end of the shallow phase, compatible with
the expectations from an external origin of the shocks, while for other GRBs the opposite is true \citep{Liang07}.

Likewise, X-ray flares have been interpreted as the result of late internal dissipation rather than
due to external shocks, one of the main arguments being their short timescales also at late times,
$\Delta t/t\ll1$ \citep{Burrows05,Falcone06,Romano06,Chincarini07,Lazzati07}, whereas an external shock origin would
require increasing timescales \citep{Fenimore96}, although see also \citet{Dermer07b}.

Another debated topic concerns the presence of optical brightening contemporaneous with the prompt
emission or occurring in the first 1--2 hours.
A variety of mechanisms have been proposed to explain it, such as
reverberation of the prompt emission radiation (GRB~050820A: Vestrand et al. 2006),\nocite{Vestrand06}
reverse shock (GRB~990123: Akerlof et al. 1999)\nocite{Akerlof99},
refreshed shocks and/or energy injection (e.g. GRB~021004: Fox et~al. 2003;\nocite{Fox03} GRB~050820A:
Cenko et al. 2006;\nocite{Cenko06} GRB~060206: Wozniak et al. 2006;\nocite{Wozniak06} Monfardini et al. 2006),
\nocite{Monfardini06}
%crossing of the wind-termination
%shock (e.g. GRB~050525A: Klotz et~al. 2005;\nocite{Klotz05} Blustin et~al. 2006), \nocite{Blustin06}
onset of the external shock (GRB~060206 and GRB~060210: Stanek et~al. 2007;
GRB~060418 and GRB~060607A: Molinari et~al. 2007),\nocite{Stanek07,Molinari07}
and large angle emission (GRB~990123: Panaitescu \& Kumar 2007).\nocite{PK07}
It is likely that the interplay of the many processes active after the GRB explosion may all affect
the optical light curves, creating the rich variety of observed behaviours.

While GRBs with complex multi-peaked time profiles displaying no pulse width evolution with
time seem to be explained more naturally through the mutual interaction of a wind of shells
emitted at different times with different Lorentz factors (internal shock model), a single FRED 
(fast rise exponential decay; Fishman \& Meegan 1995)\nocite{Fishman95} profile
can still match the expectations of a single shell sweeping up the ISM \citep{Fenimore96}.
In this scenario, the kinetic energy of a single ultrarelativistic shell is converted into
internal energy of the ISM swept up; the shocked electrons radiate via synchrotron emission
and inverse Compton scattering. In the simplest case of a thin shell ploughing into the ISM
and emitting for a short time, the expected time profile of the $\gamma$-ray prompt emission
is that of a single pulse with fast rise and slow decay.
The cooling timescale of electrons is negligible with respect to
the hydrodynamical timescale in most cases \citep{Sari97}.
As a consequence, the rise time is determined by the emission time given by the crossing time
of the shell by the reverse shock, while the decay time is dominated by the angular spreading timescale.

In this paper we report on the $\gamma$-ray, X-ray and
optical observations of GRB~070311, whose time profile is typical of a FRED.
In particular, we focus on some properties shared by the $\gamma$-ray and optical prompt
emission and the late optical and X-ray afterglow in the light of the refreshed shock scenario.

The paper is organised as follows: 
in Sect.~\ref{sec:obs} and Sect.~\ref{sec:an} we describe the observations 
and the data reduction and analysis, respectively.
Multiwavelength timing and spectral analysis
of both the prompt and the afterglow emission is presented
in Sect.~\ref{sec:multi}. In Sect.~\ref{sec:disc} we discuss our results.
Finally, in Sect.~\ref{sec:conc} we summarise our findings and conclusions.

Throughout the paper, times are given relative
to the onset time of the GRB, which corresponds to 45~s prior to
the {\rm INTEGRAL}/IBAS trigger time,
and the convention $F(\nu,t)\propto\nu^{-\beta}\,t^{-\alpha}$
has been followed, where the energy index $\beta$ is related to the
photon index $\Gamma=\beta+1$.
%We have adopted a standard cosmology model with Hubble constant
%H$_0$=70\,km\,s$^{-1}$\,Mpc$^{-1}$ and cosmological constants
%$\Omega_\Lambda=0.7$, $\Omega_M=0.3$.

All the quoted errors are given at 90\% confidence level for one interesting
parameter ($\Delta$\chisq=2.706), unless stated otherwise.

%%%%%%%%%%%%%%%%%%%%%%%%%%%%%%%%%%%%%%%%%%%%%%
\section{Observations}
\label{sec:obs}
%%%%%%%%%%%%%%%%%%%%%%%%%%%%%%%%%%%%%%%%%%%%%%

GRB~070311 triggered the {\rm INTEGRAL}/IBAS in IBIS/ISGRI data on 
2007 March 11 at 01:52:50~UT and it was localised at 
RA $= 05^{\rm h}$  $50^{\rm m}$  $09\fs86$,   
Dec. $=+03^{\circ}$ $22^{\prime}$ $29\farcs3$, 
%\RA{05}{50}{09.86}, \decl{+03}{22}{29.3} (J2000) 
 with an error radius of $2\farcm5$ \citep{Mereghetti07}.
At the time of the burst {\rm Swift}/BAT was pointing in almost the opposite direction
and it would have been in the field of view two minutes later.
The corresponding flux at the position of the burst in a 300-s image beginning
130~s after the {\rm INTEGRAL} trigger is negligible.

The $\gamma$-ray prompt emission in the 20--200~keV energy band lasted
about 50~s with a peak flux of 0.9~ph~cm$^{-2}$~s$^{-1}$ (1~s integration time)
and a fluence of $(2\div3)\times10^{-6}$~erg~cm$^{-2}$ \citep{Mereghetti07,Sazonov07}.

The Rapid Eye Mount\footnote{{\texttt http://www.rem.inaf.it/}} (REM; Zerbi et al. 2001)\nocite{Zerbi01}
telescope reacted promptly and began observing 55~s
after the GRB trigger time (see Sect.~\ref{sec:gamma}) and discovered a bright fading IR counterpart
within the {\rm INTEGRAL} error circle at 
RA $= 05^{\rm h}$  $50^{\rm m}$  $08\fs21$,   
Dec. $=+03^{\circ}$ $22^{\prime}$ $30\farcs3$ 
%\RA{05}{50}{08.21}, \decl{+03}{22}{30.3}
(J2000; Covino et~al.~2007), \nocite{Covino07} corresponding to Galactic coordinates $(l, b)$ of
$(202\fdg766$, $-11\fdg998)$.
The afterglow was soon confirmed by PAIRITEL \citep{Bloom07}.

The {\rm Swift} narrow field instruments, XRT and UVOT, began observing at 7004~s
after the trigger time.
The XRT found an uncatalogued fading source at the position 
RA $= 05^{\rm h}$  $50^{\rm m}$  $08\fs43$,   
Dec. $=+03^{\circ}$ $22^{\prime}$ $30\farcs0$
%\RA{05}{50}{08.43}, \decl{+03}{22}{30.0} 
(J2000), with an error radius of $3\farcs8$ and
$3\farcs3$ from the optical counterpart \citep{Guidorzi07a}.
No optical source was detected in correspondence of the optical and X-ray afterglows
by UVOT down to $V=19.5$ and $B=20.5$ ($3 \sigma$) from 197-s exposures \citep{Holland07}.

The 2.2--m telescope of Calar Alto (CAHA) equipped with 
the Calar Alto Faint Object Spectrograph (CAFOS) observed the afterglow
in $R$ filter at 0.72 and 1.73 days after the burst.

Because of the low Galactic latitude, the Galactic reddening along the
direction to the GRB is large: $E_{B-V}=0.763$~mag \citep{Schlegel98}.
The Galactic extinction in each filter has been estimated
through the NASA/IPAC  Extragalactic Database extinction
calculator\footnote{{\texttt http://nedwww.ipac.caltech.edu/forms/calculator.html}}. Specifically,
% from $A_V/E_{B-V}=3.31$ (appendix B of Schlegel et al. 1998),
the extinction in the other filters is derived through the parametrisation by
\citet{Cardelli89}: $A_V=2.53$, $A_R=2.04$, $A_I=1.48$, $A_J=0.69$, $A_H=0.44$, $A_K=0.28$~mag.

%%%%%%%%%%%%%%%%%%%%%%%%%%%%%%%%%%%%%%%%%%%%%%
\section{Data reduction and analysis}
\label{sec:an}
%%%%%%%%%%%%%%%%%%%%%%%%%%%%%%%%%%%%%%%%%%%%%%

%---------------------------------------------
\subsection{Gamma--ray data}
\label{sec:gamma}
%---------------------------------------------

\begin{figure}
\centering
\includegraphics[width=8.5cm]{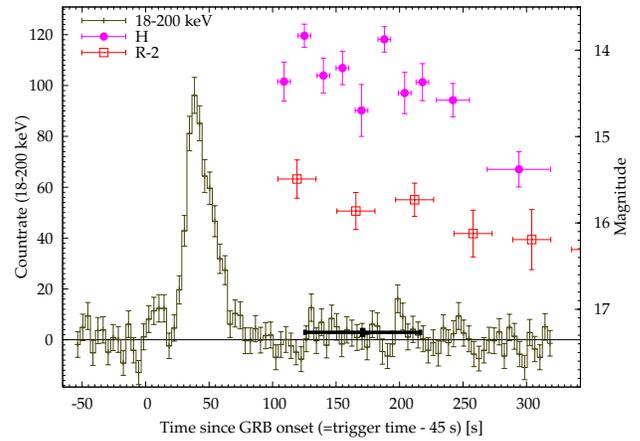}
\caption{Histogram of the {\rm INTEGRAL} 18--200~keV background-subtracted
$\gamma$-ray profile of GRB~070311 (integration time of 4~s; left axis).
The thick cross shows a grouped bin of the $\gamma$-ray profile,
$\sim2.5~\sigma$ above the background.
REM $H$ (filled circles) and $R$ (empty squares) magnitudes of the
NIR/optical afterglow are also reported on the right axis.
}
\label{fig:INT_REM}
\end{figure}

Figure~\ref{fig:INT_REM} shows the 18--200~keV
background-subtracted time profile of GRB~070311 as recorded by {\rm INTEGRAL}
\citep{Sazonov07}.
The onset of the GRB appears to occur about 45~s before the trigger time.
Hereafter, time will be referred to the onset time, i.e. 45~s prior to the
trigger time.
From 75 to $\sim125$~s the signal drops below the sensitivity of the instrument.
Interestingly, between $\sim125$ and $225$~s there is a hint of the presence of
a faint gamma-ray tail. Although this feature is detected at $\sim2.5$-$\sigma$ confidence
level (see thick cross in Fig.~\ref{fig:INT_REM}) and so should be regarded
as tentatively detected, it is interesting
to see how it fits into the overall picture together with NIR, optical and
X-ray observations of GRB~070311. We address this issue below.

The integrated spectrum in the 18--300~keV band is well fit by a power law
with a photon index of $\Gamma_\gamma=1.3\pm0.1$ and a fluence of
$(3.0\pm0.5)\times10^{-6}$~erg~cm$^{-2}$ (20--200~keV). The
spectrum shows no statistically significant high-energy cutoff.
The peak energy lies above 80~keV during the prompt emission.
There is also an indication of spectral softening
in the course of the burst, with $\Gamma_\gamma$ evolving from $0.8\pm0.2$ during
the rise phase to $1.45\pm0.15$ during the peak and decay of the emission \citep{Sazonov07}.

The conversion factor from rate to flux units in the 18--200~keV band is
$(1.2\pm0.2)\times10^{-9}$~erg~cm$^{-2}$~count$^{-1}$.

%Details are reported in \citet{Sazonov07}.

% ++++++++++++++ INTEGRAL LC ++++++++++++++
%++++++++++++++++++++++++++++++++++++++++++

%---------------------------------------------
\subsection{Infrared/optical data}
\label{sec:REM}
%---------------------------------------------
Early time optical and near infrared (NIR) data were collected using the 60-cm robotic
telescope REM located at the ESO La Silla observatory (Chile).
The REM focal instruments consist of a NIR camera (REMIR), operating in the
range 1.0--2.3~$\mu$m ($z'JHK'$), and an optical imager (REM Optical
Slitless Spectrograph, ROSS) with spectroscopic (slitless) and photometric
capabilities ($VRI$). A dichroic allows simultaneous observations at optical
and NIR wavelengths in two selected filters (for further information on REM
and its capabilities, see Covino et al. 2004\nocite{Covino04} and references therein).

REM reacted promptly to the {\rm INTEGRAL} GCN alert and began observing the
field of GRB~070311 55~s after the burst trigger (36~s after the reception
of the alert), following the event for $\sim$~1 hour.

For the first $\sim$~500 s the REMIR observations have been performed using only
the $H$ filter with increasing exposure times, then all the NIR filters have
been used in rotation. A similar observing strategy has been adopted in the
optical. $R$ band observations lasted $\sim$~1400 s for a total of 40
consecutive images. During the following $\sim$~2700 s, $VRI$ images have been
acquired in rotation but the optical transient was already below the
instrument detection limits for the $V$ and $I$ filters.

For both optical and NIR data sets, the reduction and the analysis followed standard 
procedures. The photometric 
calibration for the NIR was accomplished by applying average magnitude shifts computed 
using the 2MASS\footnote{{\texttt http://www.ipac.caltech.edu/2mass/} } catalogue.
The optical data were calibrated using instrumental 
zero points, checked with observations of standard stars in the SA95 Landolt field
\citep{Landolt92}, or with the SDSS\footnote{{\texttt http://www.sdss.org}}
in the case of the $z'$ filter.

\begin{figure*}
\centering
\includegraphics{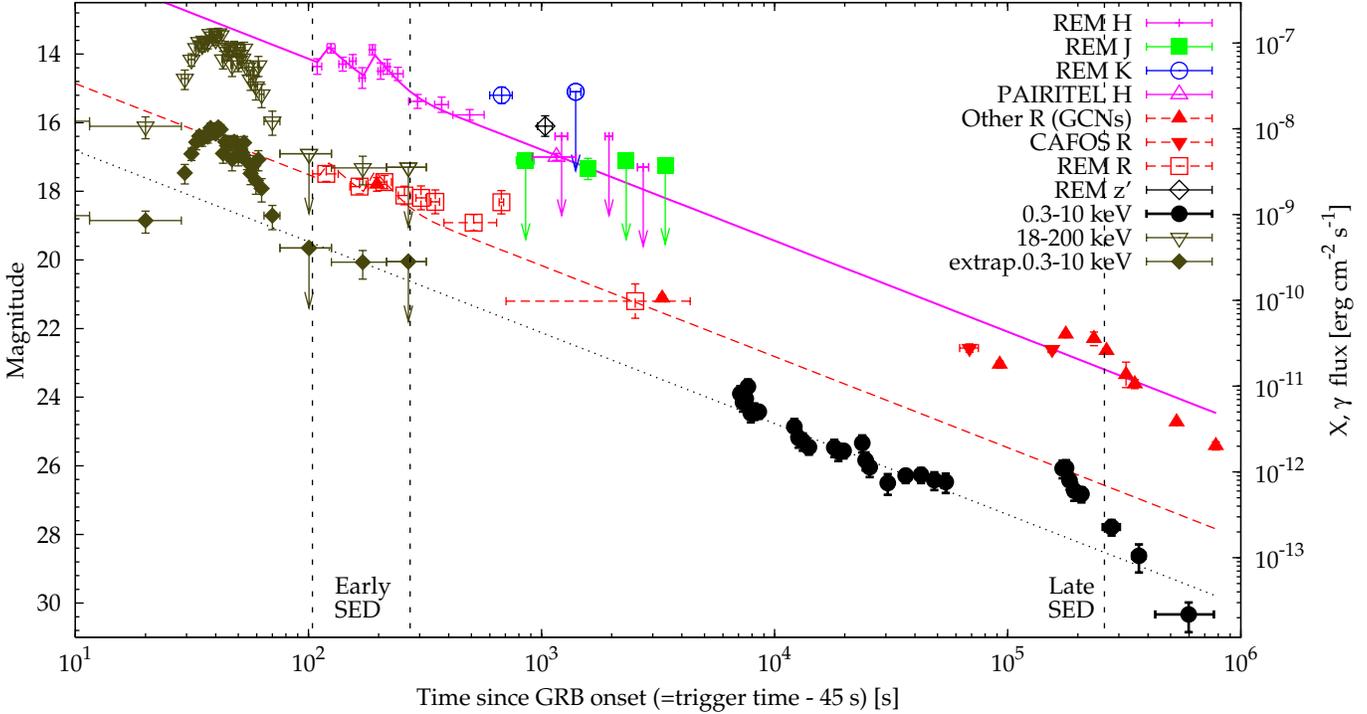}
\caption{
Panchromatic light curves of the prompt as well as of the afterglow emission of GRB~070311:
the flux in the 18--200~keV band by {\rm INTEGRAL} (empty upside down triangles)
and in the 0.3--10~keV band of the late afterglow by {\rm Swift}/XRT (filled circles) 
are given on the right axis.
Filled diamonds show the $\gamma$-ray flux extrapolated to the 0.3--10~keV band.
Magnitudes in $J$ (filled squares),
$H$ (crosses), $K$ (empty circles), $z'$ (empty diamond) and $R$ (empty squares) filters by REM
and $R$ by CAFOS (filled upside down triangles) are given on the left axis.
A single $H$ point from PAIRITEL (Bloom et~al. 2007; empty triangle) %\nocite{Bloom07}
and other $R$ points from GCN circulars
%\nocite{Cenko07,Dai07,Garnavich07,Halpern07a,Halpern07b,Halpern07c,Halpern07d,Jelinek07,Kann07,Wren07}
(Cenko 2007; Dai et~al. 2007; Garnavich et~al. 2007; Halpern \& Armstrong 2007a,b,c,d;
Jel\'inek \& Prouza 2007; Kann 2007; Wren et~al. 2007; filled triangles) are shown as well.
Upper limits are at 3-$\sigma$ confidence level.
The solid (dashed) line shows the best-fit power law ($\alpha=1.06\pm0.08$) and two pulses superposed
on the $H$ (early $R$, i.e. $t<10^4$~s) filter curve.
The dotted line shows the same power-law component renormalised to the first part ($t<10^5$~s)
of the X-ray curve ($\alpha_{\rm x}=\alpha_H=\alpha_R$).
The two earliest vertical dashed lines show the time interval used to extract an early SED
(Fig.~\ref{fig:earlysed}), while
the third vertical line shows the epoch of a late SED (Fig.~\ref{fig:latesed}). See text.
}
\label{fig:all_lc}
\end{figure*}

Figure~\ref{fig:INT_REM} shows the REM $HR$ prompt measurements (filled circles
and empty squares, respectively) together
with the $\gamma$-ray time profile, while Figure~\ref{fig:all_lc}
shows the $KHJz'R$ light curves (empty circles, crosses, filled squares, empty
diamonds, empty squares, respectively).
% All our photometric data are reported in Table~\ref{tab:REM}.

% ++++++++++++++ ALL LC ++++++++++++++
%++++++++++++++++++++++++++++++++++++++++++
%

Further observations were acquired with CAFOS.
This  instrument is  a  focal reducer  which  allows direct  imaging,
spectroscopy   and  polarimetry.    The  detector   used  is   a  SITe
2048$\times$2048 pixel CCD  providing a scale of $0\farcs53$
pixel$^{-1}$ and a circular field of view of $16\arcmin$ in diameter.
In order to reduce the CCD readout time the observations of GRB~070311
were carried  out trimming the  CCD to a 1024$\times$1024  window. The
photometric calibration  was performed observing the standard field
PG0942 \citep{Landolt92} at a similar airmass as the GRB field.
Data reduction and analysis were carried out following standard recipes
by using ESO-Eclipse (v5.0) tools \citep{Dev01}.
Aperture photometry  was obtained with the
GAIA\footnote{{\texttt http://star-www.dur.ac.uk/$\sim$pdraper/gaia/gaia.html}} (v3.2.0)
package.
REM and CAFOS photometry is reported in Table~\ref{tab:REM}.

Hereafter, the magnitudes shown are not corrected for Galactic extinction, whilst
fluxes as well as all the best-fit models are. When the models are plotted together
with magnitudes, the correction for Galactic extinction is removed from the models.

First we fitted the $H$ light curve, which is the best monitored at early times.
We tried to fit the points up to $\sim300$~s with a simple power law to test whether
the fluctuations visible in Fig.~\ref{fig:INT_REM} might be just statistical around
a power-law decay. The resulting $\chi^2/{\rm dof}$ is 17.5/7 with a chance probability
of 1.4\% and the power-law index turned out to be $\alpha_{H}=0.5\pm0.4$ (1 $\sigma$).
This probability gets even smaller if we combine the $R$ and $H$ points, as the
latter trace the former although less significantly. We infer that the probability
that the first 300~s NIR/optical profile is the result of statistical fluctuations around
a simple power-law decay is lower than 1.4\%.

This motivated us to fit the early data by means of a more detailed model.
The point from PAIRITEL at $t=1160$~s \citep{Bloom07} was included in the $H$ data set for
this analysis.
The fit shown in Fig.~\ref{fig:all_lc} (solid line) is the result of a power-law component
with decay index of $\alpha_H=1.06\pm0.08$ with two FRED-shaped
 pulses overimposed,
peaking at 119 and 180~s, respectively.
The dashed line shows the power law fitting the early part of the $R$ curve ($t<10^4$~s)
by fixing the index $\alpha_R=\alpha_H$ and letting only the normalisation free to vary.
Here we note the less pronounced enhancement in the $R$ flux at the time of the
pulses seen in $H$. A more detailed discussion of it as well as of the
fitting models is reported in Sect.~\ref{sec:multi_lc}.

For both the $K$ and $J$ magnitudes there is one single detection and the
remaining are upper limits. 
The first $J$ upper limit at 810~s preceding the detection at 1537~s is however inconsistent
with the assumption $\alpha_J=\alpha_H$ extrapolated at the time of the two measurements.
Comparing all the optical-NIR filters, a flux increase characterised by some variability
seems to appear even after the first two optical peaks, i.e. after $\sim600$~s.
This, combined with the detections in $J$ and $z'$, could be a hint of a third flare.

%%%%%%%%%%%%%%%%%%%%%%%%%%%% REM OBS %%%%%%%%%%%%%%%%%%%%%%%%%

 \begin{table*}
 \begin{center}
 \caption{REM $KHJz'R$ and CAFOS $R$ photometry of GRB~070311.}
 \label{tab:REM}
 \begin{tabular}{rrrrr}
 \hline
 \hline
 \noalign{\smallskip}
Start Time         &  End Time    &     Exposure    &   Mag$^{\mathrm{a,b}}$   &  Instr/Filter    \\
      (s)           & (s)      	& (s)       		&        		&       \\
 \noalign{\smallskip}
 \hline
 \noalign{\smallskip}
     554 &    704   &    150   & $15.20\pm0.24$ & REM/$K$ \\
    1281 &   1431   &    150   & $>$15.10       & REM/$K$ \\
\hline
      59 &     69   &     10   & $14.36\pm0.23$ & REM/$H$ \\
      75 &     85   &     10   & $13.83\pm0.14$ & REM/$H$ \\
      90 &    100   &     10   & $14.29\pm0.20$ & REM/$H$ \\
     105 &    115   &     10   & $14.21\pm0.19$ & REM/$H$ \\
     120 &    130   &     10   & $14.70\pm0.30$ & REM/$H$ \\
     138 &    148   &     10   & $13.88\pm0.15$ & REM/$H$ \\
     154 &    164   &     10   & $14.50\pm0.24$ & REM/$H$ \\
     168 &    178   &     10   & $14.37\pm0.22$ & REM/$H$ \\
     184 &    210   &     26   & $14.58\pm0.19$ & REM/$H$ \\
     224 &    274   &     50   & $15.38\pm0.20$ & REM/$H$ \\
     303 &    353   &     50   & $15.48\pm0.22$ & REM/$H$ \\
     372 &    522   &    150   & $15.77\pm0.15$ & REM/$H$ \\
    1099 &   1249   &    150   & $>$16.4  & REM/$H$ \\
    1826 &   1976   &    150   & $>$16.4  & REM/$H$ \\
    2536 &   2836   &    300   & $>$17.3  & REM/$H$ \\
\hline
     735 &    885   &    150   & $>$17.10 & REM/$J$ \\
    1462 &   1612   &    150   & $17.35\pm0.31$ & REM/$J$ \\
    2188 &   2338   &    150   & $>$17.10   & REM/$J$ \\
    3198 &   3498   &    300   & $>$17.24   & REM/$J$ \\
\hline
     917 &   1067   &    150   & $16.1\pm0.3$ & REM/$z'$ \\
\hline
    59.2 &   89.2   &     30   & $17.49\pm0.22$ & REM/$R$ \\
   105.6 &  135.6   &     30   & $17.86\pm0.21$ & REM/$R$ \\
   151.8 &  181.8   &     30   & $17.73\pm0.19$ & REM/$R$ \\
   197.9 &  227.9   &     30   & $18.12\pm0.27$ & REM/$R$ \\
   244.1 &  274.1   &     30   & $18.19\pm0.35$ & REM/$R$ \\
   290.3 &  320.3   &     30   & $18.31\pm0.35$ & REM/$R$ \\
   336.5 &  597.0   &    180   & $18.91\pm0.24$ & REM/$R$ \\
   613.0 &  643.0   &     30   & $18.32\pm0.34$ & REM/$R$ \\
   659.2 & 4295.6   &   1530   & $21.20\pm0.50$ & REM/$R$ \\
   62174 & 74783    &   8400   & $22.57\pm0.12$ & CAFOS/$R$ \\
  149180 & 160924   &  10810   & $22.60\pm0.08$ & CAFOS/$R$ \\
 \noalign{\smallskip}
  \hline
  \end{tabular}
  \end{center}
  \begin{list}{}{} 
  \item[$^{\mathrm{a}}$] Values are not corrected for Galactic extinction.
  \item[$^{\mathrm{b}}$] Errors at the 68\% confidence level and
                  upper limits (3~$\sigma$) are given. 
  \end{list}   
\end{table*}

%%%%%%%%%%%%%%%%%%%%%%%%%%%%%%%%%%%%%%%%%%%%%%%%%%%%%%%%%%%%%%%%%%%%%

%---------------------------------------------
\subsection{X--ray data}
\label{sec:X}
%---------------------------------------------
The XRT began observing GRB~070311 on 2007 March 11 at 03:49:34~UT,
7004~s after the {\rm INTEGRAL} trigger and ended on 2007 March 19
at 22:38:54~UT, with a total net exposure of 81.8~ks in photon counting (PC)
mode spread over 8.8 days. Table~\ref{tab:logXRT} reports the log of
the XRT observations.

%%%%%%%%%%%%%%%%%%%%%%%%%%%% XRT LOG OBS %%%%%%%%%%%%%%%%%%%%%%%%%%%%%%%%%%%%%%%%%%%%%

\begin{table*}
 \begin{center}
 \caption{XRT observation log of GRB~070311.}
 \label{tab:logXRT}
 \begin{tabular}{ccccrrr}
 \hline
 \hline
 \noalign{\smallskip}
Sequence & Obs Mode & Start Time (UT) &  End Time (UT) & Exposure & Start Time$^{\mathrm{a}}$ & 
            End Time$^{\mathrm{a}}$ \\
         &          &                 &       &   (s)    & (s)   &    (s)\\ 
 \noalign{\smallskip}
 \hline
 \noalign{\smallskip}
00020052001 &  PC & 2007-03-11 03:49:34 &  2007-03-11 17:08:03  &18780 &   7004 &  54913\\
00020052002 &  PC & 2007-03-13 00:49:40 &  2007-03-13 13:46:56  & 9697 & 169010 & 215646\\
00020052003 &  PC & 2007-03-14 00:49:36 &  2007-03-14 13:53:56  & 9465 & 255406 & 302466\\
00020052004 &  PC & 2007-03-15 00:47:36 &  2007-03-15 14:01:54  & 9954 & 341686 & 389344\\
00020052005 &  PC & 2007-03-16 01:06:34 &  2007-03-16 23:41:55  & 9028 & 429224 & 510545\\
00020052006 &  PC & 2007-03-17 01:08:59 &  2007-03-17 22:21:55  &14574 & 515769 & 592145\\
00020052007 &  PC & 2007-03-17 23:45:20 &  2007-03-18 22:32:56  & 6283 & 597150 & 679206\\
00020052008 &  PC & 2007-03-19 00:04:45 &  2007-03-19 22:38:54  & 3711 & 684715 & 765964\\
 \noalign{\smallskip}
  \hline
  \end{tabular}
  \end{center}
  \begin{list}{}{} 
  \item[$^{\mathrm{a}}$] Since {\rm INTEGRAL} trigger time.
  \end{list}   
\end{table*}

%%%%%%%%%%%%%%%%%%%%%%%%%%%%%%%%%%%%%%%%%%%%%%%%%%%%%%%%%%%%%%%%%%%%%%%%%%%%%%%%%%%%%%

The XRT data were processed using the FTOOLS software package (v.~6.1)
distributed within HEASOFT. We ran the task {\em xrtpipeline}
(v.0.10.4) applying calibration and standard filtering and screening
criteria. Data were acquired only in PC mode due to the faintness of the source.
Events with grades 0--12 were selected.
The XRT analysis has been performed in the 0.3--10 keV energy band.

% ++++++++++++++ XIMAGE ++++++++++++++
\begin{figure}
\centering
\includegraphics[width=8.5cm]{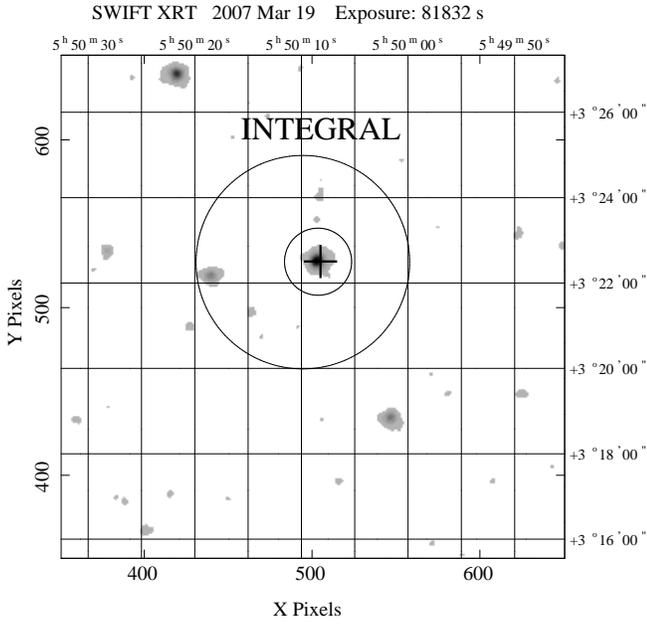}
\caption{XRT image of the field of GRB~070311 obtained from 82~ks PC data.
The large circle shows the {\rm INTEGRAL} position with an error radius
of $2\farcm5$ (90\% CL). The cross shows the optical afterglow position
discovered by REM. The small circle is a 20-pixel radius region
centred on the XRT afterglow.}
\label{fig:ximage}
\end{figure}
%++++++++++++++++++++++++++++++++++++++++++

%-----------------------------------------
\subsubsection{Temporal analysis}
\label{sec:X_lc}
%-----------------------------------------
Source photons were extracted from a circular region with a 20--pixel radius
(1~pixel$\mbox{}=2\farcs36$; Fig.~\ref{fig:ximage}) and PSF-renormalised.
The background was estimated from a four-circle region with a total area of
$11.5\times10^3$~pixel$^{2}$ away from any source present in the field.
When the count rate dropped below $\sim10^{-2}$ counts~s$^{-1}$, we made
use of {\tt XIMAGE} with the tool {\tt SOSTA},
which corrects for vignetting,
exposure variations and PSF losses within an optimised box, using the same
background region.

\begin{figure}
\centering
\includegraphics[width=8.5cm]{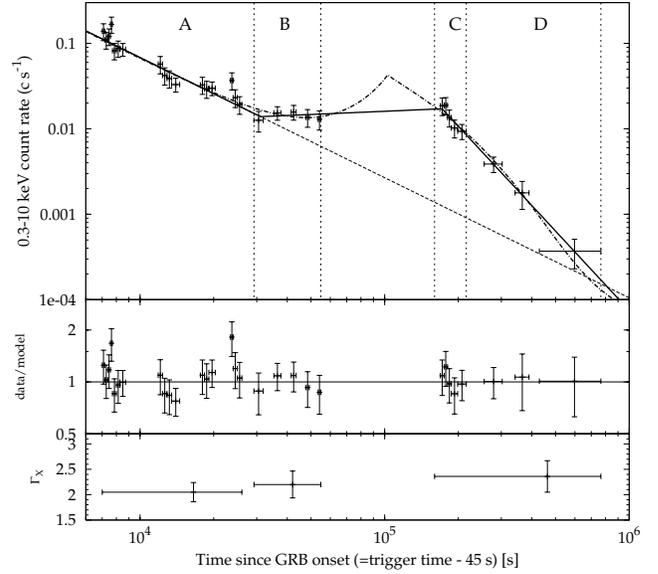}
\caption{{\em Top panel}: X-ray afterglow light curve of GRB~070311 obtained with XRT
in the 0.3--10~keV energy band. The solid line shows the best fit obtained
with a double broken power law, while the dashed line shows the extrapolation
of the initial power-law decay. The dashed-dotted line shows the best-fit combination
of a FRED-like pulse superposed on a power law. Labelled vertical slices correspond
to four different regions where spectra have been extracted.
{\em Mid panel}: fractional residuals with respect to the double broken power-law
model.
{\em Bottom panel}: photon index (error bars are 1~$\sigma$).}
\label{fig:xrt_lc}
\end{figure}

The resulting 0.3--10~keV light curve is shown in Fig.~\ref{fig:xrt_lc}.
It was binned so as to achieve a minimum signal to noise ratio (SNR) of 3,
except for the two last bins (with a SNR of 2.8 and 2.6, respectively)
as well as a minimum of 20 total counts.

% ++++++++++++++ XRT LC ++++++++++++++
%++++++++++++++++++++++++++++++++++++++++++

The light curve has been fit firstly with a double broken power law (solid line in
Fig.~\ref{fig:xrt_lc}), whose best fit parameters are the following:
$\alpha_{\rm x,1}=1.4\pm0.1$, $\alpha_{\rm x,2}=-0.1_{-0.2}^{+0.7}$,
$\alpha_{\rm x,3}=3.1_{-0.4}^{+0.5}$,
$t_{\rm x,b1}=31\pm4$~ks, $t_{\rm x,b2}=1.7_{-0.4}^{+0.1}\times10^5$~s
($\chi^2/{\rm dof}=16.8/24$), where $\alpha_{{\rm x},i}$ ($i=1,2,3$) are the
canonical initial, mid, and final decay slopes
and $t_{{\rm x,b}i}$ ($i=1,2$) are the two break times, respectively.
The last point lies on the extrapolation of the initial power-law decay
(dashed line in Fig.~\ref{fig:xrt_lc}).
From the X-ray data alone it is not possible to determine whether
the shallow and final steep decay phases are the result of a late rebrightening
after which the decay will resume to the pre-break behaviour.

\begin{figure}
\centering
\includegraphics[width=8.5cm]{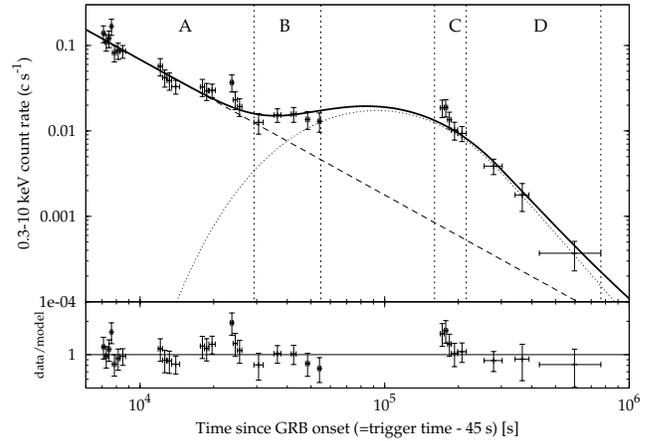}
\caption{{\em Top panel}: X-ray afterglow light curve of GRB~070311 obtained with XRT
in the 0.3--10~keV energy band. The curve has been fit with the
two-component model by \citet{Willingale07}: the prompt (dashed),
the afterglow (dotted) and their combination (solid). Labelled intervals are the
same as in Fig.~\ref{fig:xrt_lc}.  
{\em Bottom panel}: fractional residuals with respect to the model.}
\label{fig:xrt_willingale}
\end{figure}

We also adopted two alternative models: the combination of a FRED-shaped
pulse with a power law (dash-dotted line in Fig.~\ref{fig:xrt_lc}; a more detailed
description follows in Sect.~\ref{sec:pl_pulses}), and the model by
\citet{Willingale07}. This model is
the combination of two components, the prompt and the afterglow (according to
the terminology introduced by these authors), described with
the same functional form, which consists of a combination of an exponential and
a power law. Following the notation by \citet{Willingale07}, we fixed the rise time
of the prompt component to the onset time: $t_{\rm p}=0$.
We also fixed the time when the power-law component of the prompt takes over,
$T_{\rm p}=5\times10^3$~s, i.e. prior to the beginning of the X-ray observations,
when the X-ray decay is already dominated by the power law.
The best-fit model is shown in Fig.~\ref{fig:xrt_willingale}.
% ++++++++++++++ XRT LC ++++++++++++++
%++++++++++++++++++++++++++++++++++++++++++
%
The best-fit parameters are the following: $\alpha_{\rm p}=1.6\pm0.2$,
$t_{\rm a}=(1.0\pm0.4)\times10^5$~s, $T_{\rm a}=(2.9\pm1.0)\times10^5$~s,
$\alpha_{\rm a}=3.5\pm0.7$ ($\chi^2/{\rm dof}=24.9/24$).
We point out that within this model the late rebrightening corresponds
to the onset of the afterglow component: this forced us to decouple $t_{\rm a}$
from $T_{\rm p}$ and treat the former as a free parameter,
unlike what \citet{Willingale07} did for all of the GRBs of their sample.
We last note that the two-component model by \citet{Willingale07}
usually accounts for both the prompt and the afterglow emission. In our
fit, we have only considered the late-time light curve. The X-ray
rebrightening observed in GRB~070311 would therefore constitute a third
component, which is only rarely seen in afterglow light curves.

%-----------------------------------------
\subsubsection{Spectral analysis}
\label{sec:X_spec}
%-----------------------------------------
In order to detect possible spectral variations connected with changes in
the X-ray light curve, we extracted the 0.3--10~keV spectrum in four
different time intervals, labelled ``A'', ``B'', ``C'' and ``D'',
corresponding to the initial steep decay, the beginning
of the shallow phase, the end of it (or the peak of the rebrightening, according
to the alternative description of the X-ray light curve discussed in Sect.~\ref{sec:X_lc}),
and the final steep decay, respectively (see Fig.~\ref{fig:xrt_lc}).
%
% ++++++++++++++ XRT spectra ++++++++++++++
\begin{figure}
\centering
\includegraphics[angle=270,width=8.5cm]{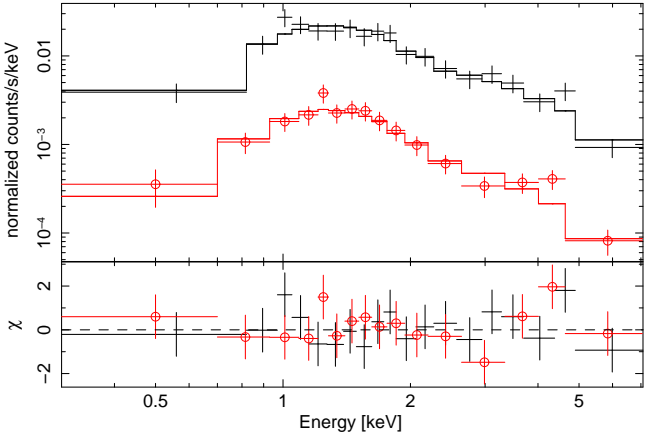}
\caption{{\em Top Panel}: X-ray afterglow photon spectra of GRB~070311
obtained with XRT in the 0.3--10~keV energy band corresponding to the temporal
interval A (crosses) and BCD (circles), respectively.
Solid lines show the corresponding best-fit absorbed power laws.
{\em Bottom Panel}: residuals with respect to the corresponding best-fit models.}
\label{fig:xrt_spec}
\end{figure}
%++++++++++++++++++++++++++++++++++++++++++
%

Source and background spectra were extracted from the same regions as the ones
used for the light curve (Sect.~\ref{sec:X_lc}), except for intervals ``B'' and
``C'' separately. For these intervals, due to their poor statistics, it was not possible to group the energy
channels so as to have a Gaussian distribution of the number of photons per grouped channel.
Hence, for both spectra we replaced the $\chi^2$ with the $C$ statistics \citep{Cash79},
which has proven to be useful whenever the Gaussian approximation does not hold (e.g. when the number
of photons per bin is less than 10), provided that the contamination of background photons
is negligible. In order to ensure this, for both spectra ``B'' and ``C'',
a 10--pixel radius circular region was used.

The ancillary response files were generated using the task {\tt xrtmkarf}.
Spectral channels were grouped so as to have at least 20 counts per bin, except for ``B'' and ``C''.
Spectral fitting was performed with {\tt xspec} (v.~11.3.2).

%%%%%%%%%%%%%%%%%%%%%%%%%%% XRT Spectra %%%%%%%%%%%%%%%%%%%%%%%%%%%%%%%%%%%%%%%%%%%%%
 \begin{table*}
 \begin{center}
 \caption{Best-fit parameters of the 0.3--10~keV spectrum
of the X-ray afterglow. The model is an absorbed power law ({\tt xspec} model: {\sc wabs}
{\sc pow}).}
 \label{tab:xspec}
 \begin{tabular}{lrrcccc}
 \hline
 \hline
 \noalign{\smallskip}
 Interval  & Start time  & Stop time  & $N_{\rm H}$  & $\Gamma_{\rm x}$  & Mean flux  & $\chi^2$/dof  \\
          & (s)    & (s)  & ($10^{21}$~cm$^{-2}$)  &   & ($10^{-13}$~erg~cm$^{-2}$~s$^{-1}$)  &    \\
 \noalign{\smallskip}
 \hline
 \noalign{\smallskip}
A  &   7004 &  26079 & $4.5_{-1.0}^{+1.3}$ & $2.05_{-0.24}^{+0.27}$ & $24\pm4$    & 11.0/16\\
B  &  29217 &  54833 & $4.2_{-2.1}^{+3.2}$ & $2.2_{-0.5}^{+0.6}$    & $6.4\pm1.9$ &  351.5 (63.4\%)$^{a}$\\
C  & 160010 & 215648 & $3.4_{-1.4}^{+1.7}$ & $2.2_{-0.4}^{+0.5}$    & $5.6\pm1.4$ &  320.4 (58.2\%)$^{a}$\\
BC &  29217 & 215648 & $5.0_{-1.9}^{+1.9}$ & $2.5_{-0.5}^{+0.6}$    & $4.5\pm1.5$ & 12.4/9\\
BCD&  29217 & 765966 & $5.5_{-1.7}^{+2.1}$ & $2.5\pm0.2$            & $1.9\pm0.1$ & 10.3/13\\
CD & 160010 & 765966 & $4.4_{-1.3}^{+2.1}$ & $2.4\pm0.4$            & $1.1\pm0.4$ &  6.2/7\\
%2  &  29217 &  54833 & $9.5_{-4.0}^{+6.9}$ & $3.2_{-0.8}^{+1.1}$    & $5.4\pm1.4$ &  1.9/3\\
%3  & 160010 & 215648 & $4.7_{-2.3}^{+3.1}$ & $2.9_{-0.8}^{+0.9}$    & $4.0\pm1.0$ &  0.14/2\\
 \noalign{\smallskip}
  \hline
  \end{tabular}
  \end{center} 
\flushleft
$^{a}$ Cash statistics ($C$-stat; Cash 1979)\nocite{Cash79} and percentage of Monte Carlo realisations
that had statistic $<$ $C$-stat. We performed $10^4$ simulations. Photons were extracted from a circular
region with a 10--pixel radius.\\
\end{table*}

%%%%%%%%%%%%%%%%%%%%%%%%%%%%%%%%%%%%%%%%%%%%%%%%%%%%%%%%%%%%%%%%%%%%%%%%%%%%%%%%%

All the spectra can be modelled with an absorbed power law with the {\tt xspec} models
{\sc wabs pow}.
We assumed the photoelectric cross section by \citet{Morrison83}.
Results of the best-fit parameters are reported in Table~\ref{tab:xspec}.

The Galactic neutral Hydrogen column density along the GRB direction from
21-cm-line radio surveys is $N_{\rm HI}^{\rm (Gal)}=2.3\times10^{21}$~cm$^{-2}$
\citep{Kalberla05}. The X-ray absorption found from spectral fitting is significantly
higher than $N_{\rm HI}^{\rm (Gal)}$: specifically, it is about twice as high.
We do not interpret this as evidence for intrinsic absorption:
for low Galactic latitudes ($b<25^{\circ}$) the $N_{\rm H}$ measured from X rays
is about twice as high as that derived from the radio, interpreted as due to the
presence of molecular gas \citep{Arabadjis99,Baumgartner06}.
Therefore we conclude that the $N_{\rm H}$ we measure from the X-ray spectrum is
consistent with the Galactic value expected in the direction of GRB~070311, albeit
we cannot exclude some intrinsic absorption.

Comparing the best-fit parameters obtained for the different spectra, while the
absorption seems constant, we find the suggestion ($2.2 \sigma$)
% Photon index 1 with 1-sigma unc: 2.05 +/- 0.19
% Photon index 2 with 1-sigma unc: 2.51 +/- 0.074
for a softening of the photon index
$\Gamma_{\rm x}$, from $2.05_{-0.24}^{+0.27}$ (interval A) to $2.5\pm0.2$
(intervals B, C, D merged together). The corresponding spectra with the best-fit
models are shown in Fig.~\ref{fig:xrt_spec}.

We note that the change of the spectral index, $\Delta\beta_{\rm x}=\Delta\Gamma_{\rm x}$,
is consistent with the canonical value of $1/2$ expected in the standard synchrotron
model when the cooling frequency crosses the observed passband (X rays, in this case).
In this case, this passage would have occurred approximately between intervals
A and BCD (Fig.~\ref{fig:xrt_lc}).

The excess visible between 4 and 5~keV in the residuals of the spectrum with
respect to the absorbed power law appears to be $\sim2.5$-$\sigma$ significant after
properly rebinning and in the most favourable case (Fig.~\ref{fig:xrt_spec}).

%%%%%%%%%%%%%%%%%%%%%%%%%%%%%%%%%%%%%%%%%%%%%%
\section{Multi-band combined analysis}
\label{sec:multi}
%%%%%%%%%%%%%%%%%%%%%%%%%%%%%%%%%%%%%%%%%%%%%%

%---------------------------------------------
\subsection{Light curves fitting}
\label{sec:multi_lc}
\begin{figure}
\centering
\includegraphics[width=8.5cm]{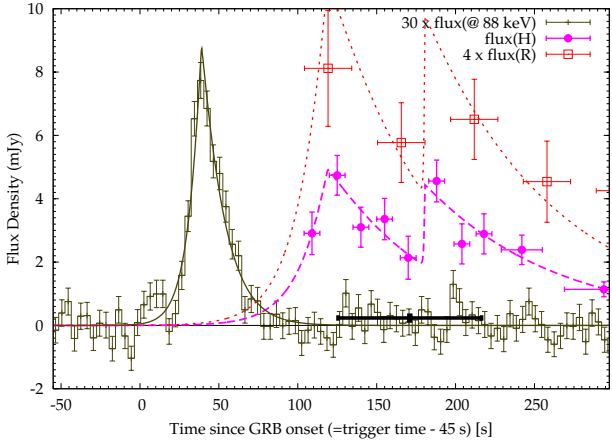}
\caption{{\rm INTEGRAL} background-subtracted
$\gamma$-ray flux density at 88~keV (crosses; integration time of 4~s) magnified 30 times,
REM $H$ (filled circles) and $R$ (empty squares; magnified 4 times) flux densities of the
NIR/optical afterglow (corrected for Galactic extinction).
Solid, dashed and dotted lines show the best-fit models of the $\gamma$-ray, $H$-
and $R$-band pulses, respectively. The thick cross is the same as that in Fig.~\ref{fig:INT_REM}.}
\label{fig:all_pulses}
\end{figure}

%---------------------------------------------
We analysed the different decays in different energy bands as homogeneously
as possible. Motivated by the possible interpretation of the FRED in the external
shock context, and by the shape of the $H$-filter pulses similar to that
of the prompt emission in the 18--200~keV band, which looks like a typical
FRED, we first fitted the 18--200~keV pulse adopting the profile used by \citet{Norris96}:
\begin{eqnarray}
\displaystyle F(t) =
\displaystyle \left\{\begin{array}{l}
\displaystyle A\ \exp{\Big[-\Big(\frac{t_{\rm max}-t}{\sigma_{\rm r}}\Big)^\nu\Big]}\quad,\quad t<t_{\rm max}\\
\displaystyle A\ \exp{\Big[-\Big(\frac{t-t_{\rm max}}{\sigma_{\rm d}}\Big)^\nu\Big]}\quad,\quad t>t_{\rm max}\\
\end{array}
\right.
\label{eq:norris}
\end{eqnarray}
where $t_{\rm max}$ is the peak time, $\sigma_{\rm r}$ and $\sigma_{\rm d}$ are the rise and
decay times, respectively, $A$ is the normalisation and $\nu$ is the peakedness (when $\nu=1$
the profile is a simple exponential, when $\nu=2$ it is a Gaussian).
The best-fit parameters for the $\gamma$-ray pulse are: $t_{\rm peak}=39.0\pm0.8$~s,
$\sigma_{\rm r}=8.5\pm1.0$~s, $\sigma_{\rm d}=15.0\pm1.7$~s. The peakedness was found
to be $\nu=1.08\pm0.13$ starting with an initial value of $\nu=1$.
The result is shown by the solid line in Fig.~\ref{fig:all_pulses}.
%
% ++++++++++++++ ALL PULSES ++++++++++++++
%++++++++++++++++++++++++++++++++++++++++++
%
We calculated the flux density at 88~keV, which corresponds to the energy
at which the flux density equals the average flux density in the 18--200~keV band
assuming $\beta_\gamma=0.3$.

To fit the curves, we adopted two alternative approaches.
First we focused on the early pulses and fitted them in terms of two overlapping
FRED-shaped pulses ({\sc pulses} model).
We chose to model the shape of the pulses with a FRED, first because this
fits well, and, secondly, because it allows a better comparison with the
results of the fit of the $\gamma$-ray pulse.
Alternatively, we added a power-law continuum, in the assumption
that the afterglow contribution is not negligible at this time
({\sc pl+pulses} model).
Both $H$ and $R$ profiles have been corrected for Galactic extinction.
Given the less dense sampling of the $H$ and $R$ curve with respect to the $\gamma$-ray
one, the peakedness was fixed to the best-fit value reported above.

Finally we fitted the late rebrightening seen in $R$ and X-ray with a
single pulse superposed to a power law ({\sc pl+pulse} model).
Best-fit parameters of the models are reported in Table~\ref{tab:multi_lc}.

%---------------------------------------------
\subsubsection{{\sc pulses} model}
\label{sec:pulses}
%---------------------------------------------
The first $H$ pulse peaks 80~s after the $\gamma$-ray peak. The pulse shape is
different: the rise time of the $H$ pulse is about twice as long,
while its decay time is about 4 times longer.
The flux density normalisation constant, $A_{\rm H1}=4.9\pm0.3$~mJy, is
about 17 times that of the $\gamma$ rays at 88~keV.
The second $H$ pulse follows the first by $\sim62$~s. Because of the poor sampling
of the rise, only the decay is better constrained and turns out to be about twice
as long as the decay time of the first pulse, while the normalisation of the second
pulse is about half as big: $A_{\rm H2}\simeq A_{\rm H1}/2$,
so that the fluence during the decay is similar to that of the first pulse.
The dashed line in Fig.~\ref{fig:all_pulses} shows
the sum of both pulses fitting the $H$ points (filled circles).

% ++++++++++++++ ALL PULSES ++++++++++++++
\begin{figure}
\centering
\includegraphics[width=8.5cm]{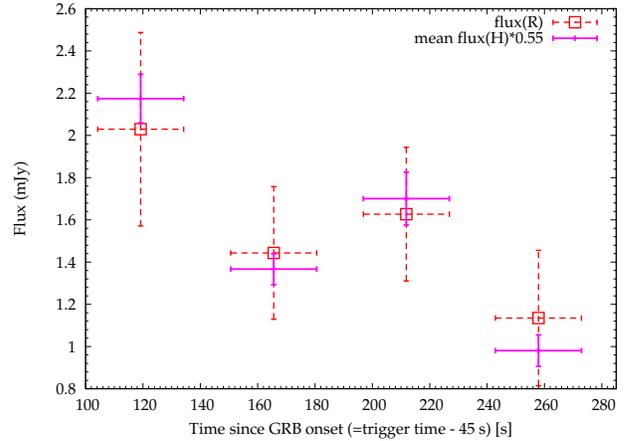}
\caption{Early $R$ optical flux (dashed squares) compared with the time-integrated
flux expected from the best-fit model of the $H$ curve rescaled by a factor
of $0.55\pm0.06$, which minimises the $\chi^2$.}
\label{fig:R_vs_integrated_H}
\end{figure}
%++++++++++++++++++++++++++++++++++++++++++
%
To evaluate whether the $R$ points simultaneous to the prominent pulses
seen in $H$ are consistent with being derived from the same profile as $H$
within uncertainties, we integrated the best-fit model of the $H$ curve
over the $R$ time bins. We determined the rescaling
factor $f_{\rm best}$ that minimises the $\chi^2$ between the measured $R$ and the
expected rescaled mean $H$ flux. The result is shown in Fig.~\ref{fig:R_vs_integrated_H}
and corresponds to $f_{\rm best}=0.55\pm0.06$, yielding $\chi^2/{\rm dof}=0.36/3$.
We conclude that the time profile best-fitting the $H$ pulses is consistent with the
simultaneous $R$ measurements.

$R$ points are displayed in Fig.~\ref{fig:all_pulses} with empty squares and the best-fit
model is represented with the dotted line, all magnified by a factor of 4 for the sake
of clarity.

%---------------------------------------------
\subsubsection{{\sc pl+pulses} model}
\label{sec:pl_pulses}
%---------------------------------------------
Alternatively to the description of Sect.~\ref{sec:pulses}, here we assume that the contribution
of the power-law continuum is not negligible since the very beginning of the NIR/optical observations.
The result is shown in Fig.~\ref{fig:all_lc}.
For the same reasons as in Sect.~\ref{sec:pulses} we first fitted the $H$ profile.
The power-law continuum turned out to have a slope of $\alpha_H=1.06\pm0.08$ (solid line
in Fig.~\ref{fig:all_lc}). Then we added the same combination of pulses as that found
in Sect.~\ref{sec:pulses}, by allowing the single normalisations and releasing gradually
some of the parameters. Table~\ref{tab:multi_lc} reports the best-fit values.
Similarly to Sect.~\ref{sec:pulses}, the fit for the $R$ filter was done by allowing just a scaling
factor of the $H$ profile. This turned out to be $f_{\rm best}=0.55\pm0.07$, i.e. the same
as that obtained with the previous model (Sect.~\ref{sec:pulses}).
The $R$ best-fit profile is shown in Fig.~\ref{fig:all_lc} with a dashed line.
The $\chi^2/{\rm dof}$ of the best-fit model are acceptable: 6.0/7 and 12.1/8 for the
$H$ and $R$ profiles, respectively.

The main differences from the results obtained in Sect.~\ref{sec:pulses} concern the shorter rise
times in this case, that were fixed to 0.1~s. This is a consequence of having increased the
continuum component here represented by the power law.
The pulse shape is less constrained, for the same reason. The second pulse still has a longer
decay than the first, while its peak intensity is now comparable with that of the first.

The dotted line in Fig.~\ref{fig:all_lc} shows the best-fit power law of the X-ray data
up to $10^5$~s obtained by fixing $\alpha_{\rm x}=\alpha_H=\alpha_R$. The fit is
acceptable: $\chi^2/{\rm dof}=24.3/21$ (chance probability of 28\%). This description of the X-ray
light curve is alternative to that given in Sect.~\ref{sec:X_lc}.

We tentatively extrapolated the $\gamma$-ray flux to the 0.3--10~keV band, assuming the
simple power law with $\Gamma_\gamma=1.3\pm0.1$ (Sect.~\ref{sec:gamma}). This assumption relies upon
the fact that the peak energy is likely to lie above 80~keV because of the hardness of the
photon index $\Gamma_\gamma$. We also applied the suppression factor due to the soft X-ray
absorption measured in the X-ray spectrum. The result is shown with filled diamonds in
Fig.~\ref{fig:all_lc}. Interestingly the single power law derived from fitting the early
$H$ and $R$ continuum components, $\alpha_{\rm x}=1.06$, matches the level of the
extrapolated prompt emission flux in the X-ray band.

%---------------------------------------------
\subsubsection{Late $R$/X-ray rebrightening}
\label{sec:late_pl_pulse}
\begin{figure}
\centering
\includegraphics[width=8.5cm]{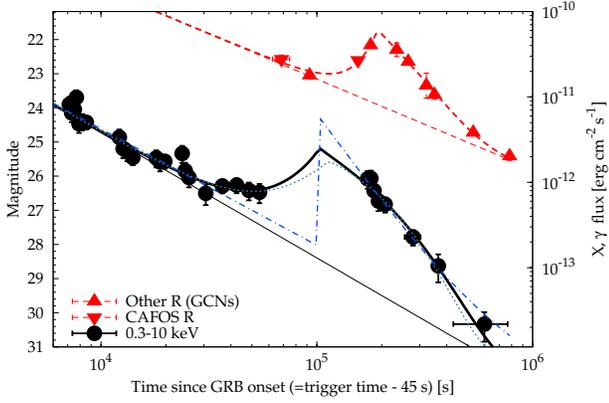}
\caption{Late time rebrightening and correspondent best-fit models
(power law and a pulse) seen in $R$ filter (triangles, thick dashed line)
and 0.3--10~keV band (circles, thick solid line). Thin lines show the
correspondent power-law component alone. X-rays: the dotted line shows the rescaled version
of the $\gamma$-ray prompt pulse combined with the underlying power law, while
the dashed-dotted line represents the sum of two power laws.
}
\label{fig:RX_bumps}
\end{figure}

%---------------------------------------------
From Fig.~\ref{fig:all_lc} we note that the power-law continuum, over which the late
($t\sim2\times10^5$~s) $R$ rebrightening sets in, looks brighter than expected from
extrapolating the early best-fit model by $\sim$~1 order of magnitude. 
For this reason we fitted it separately from the early $R$ light curve.
We adopted a combination of a power-law and a single pulse (Eq.~\ref{eq:norris}).
The power-law index was fixed to the value found in Sect.~\ref{sec:pl_pulses}
for the early part of the $R$, i.e. $\alpha_R=1.06$.
The result is shown in Fig.~\ref{fig:RX_bumps} (thick dashed line).
%
% ++++++++++++++ LATE REBRIGHTENING: R and X-ray ++++++++++++++
%++++++++++++++++++++++++++++++++++++++++++
%
We did the same for the contemporaneous 0.3--10~keV profile (thick solid line in
Fig.~\ref{fig:RX_bumps}). We initially fixed
$\alpha_{\rm x}=\alpha_R$ in analogy with Sect.~\ref{sec:pl_pulses}, but we had to
release it because of the badness of the fit and it turned out to be
$\alpha_{\rm x}=1.47\pm0.20$, more consistent with the broken-power-law fit given
in Sect.~\ref{sec:X_lc}. The X-ray coverage of the late
rebrightening is not as detailed in catching the peak time as for the $R$ filter,
nonetheless from Fig.~\ref{fig:RX_bumps} we can confidently affirm that X-rays peak
earlier than optical, which is rising or flattening during the beginning of the
monitored X-ray decay.
Best-fit parameters are reported at the bottom of Table~\ref{tab:multi_lc}.

The dashed-dotted line in Fig.~\ref{fig:RX_bumps} represents the case when the
X-ray late rebrightening is fit with the sum of two power laws,
the second of which turns on between $6\times10^4$ and $2\times10^5$~s and
models the rebrightening superposed to the first power law.
The best-fit indices are $\alpha_{\rm x,1}^{\rm (pp)}=1.4\pm0.1$ and
$\alpha_{\rm x,2}^{\rm (pp)}=3.5\pm0.5$, respectively (1~$\sigma$ confidence).
For both components the time origin was fixed to the GRB onset time.
The time at which the second power law sets in cannot be estimated from our
data; however, this is irrelevant for determining the power-law indices.

Compared to the $R$ band, the poorer coverage of the X-ray peak reflects upon
bigger uncertainties on the best-fit parameters of the X-ray profile.
Interestingly, rise and decay times are similar and the ratios
$\sigma_{\rm d}/\sigma_{\rm r}$ are
2.3 and 2.6 for the $R$ filter and X-ray, respectively, i.e. the pulses resemble the
typical shape of a prompt $\gamma$-ray pulse \citep{Norris96}.
The peak intensities are $6.5\pm0.9$ and $18\pm9$ times the value of the correspondent
underlying power-law at peak time for the $R$ filter and X-ray band, respectively.
The X-ray/$R$ time lag amounts to $(9.1\pm3.3)\times10^4$~s. More simply, the late $R$
rebrightening peaks twice as late as the X-ray profile. A word of caution is
needed about the evaluation of this temporal lag: the uncertainty
could be larger than our estimate, which is constrained by the choice of the functional
form used for fitting. Nevertheless, the evidence for a positive lag is
apparent, regardless of the fits. 

%---------------------------------------------
\subsubsection{Prompt $\gamma$ pulse/late X-ray rebrightening}
\label{sec:gamma_late_pl_pulse}
%---------------------------------------------
Notably, the ratios between the temporal parameters best fitting the pulse of
the $\gamma$-ray prompt profile (Sect.~\ref{sec:pulses}) and their correspondent
ones fitting the late X-ray rebrightening (Sect.~\ref{sec:late_pl_pulse})
are all comparable: $t_{\rm peak,late X}/t_{{\rm peak},\gamma}=2600\pm800$,
$\sigma_{\rm r,late X}/\sigma_{{\rm r},\gamma}=3400\pm3000$,
$\sigma_{\rm d,late X}/\sigma_{{\rm d},\gamma}=4400\pm800$.
We tried to fit the late X-ray rebrightening with the combination of the same
power law as that obtained in Sect.~\ref{sec:late_pl_pulse} and a rescaled version
of the $\gamma$-ray prompt pulse:
$t_{\rm peak,late X}=f_{\rm s,X}\,(t_{{\rm peak},\gamma}-t_0)$,
$\sigma_{\rm r,late X}=f_{\rm s,X}\,\sigma_{{\rm r},\gamma}$,
$\sigma_{\rm d,late X}=f_{\rm s,X}\,\sigma_{{\rm d},\gamma}$.
We left three parameters free to vary: the scaling factor $f_{\rm s,X}$,
the time origin $t_0$ and the normalisation of the pulse.
The choice of letting the time origin vary was motivated by the peak
time of the $\gamma$-ray pulse being very sensitive due to its smallness.
The result is shown in Fig.~\ref{fig:RX_bumps} (dotted line).
The best-fit parameters are the following:
$f_{\rm s,X}=5700\pm700$ and $t_0=-19.0\pm2.4$~s ($\chi^2/{\rm dof}=19.3/27$).
The potentially strong implications on the interpretation of this
result are addressed in Sect.~\ref{sec:disc}.

%%%%%%%%%%%%%%%%%%%%%%%%%%% MULTI-LC FITTING %%%%%%%%%%%%%%%%%%%%%%%%%%%%%%%%%%%%%%%%%%%%%
 \begin{table*}
 \begin{center}
 \caption{Best-fit parameters of the multi-band light curves
of the afterglow of GRB~070311. Uncertainties are 1~$\sigma$. Values of frozen parameters are
reported among square brackets.}
 \label{tab:multi_lc}
 \begin{tabular}{ccccccccc}
 \hline
 \hline
 \noalign{\smallskip}
Energy  & Model  & Component   & $\alpha$  &
$t_{\rm peak}^{\mathrm{a}}$  & $\sigma_{\rm r}$  & $\sigma_{\rm d}$  & $A$ & $\nu$  \\
band/filter          &      &   &   & (s)      & (s)  & (s)  & (mJy)  &     \\
 \noalign{\smallskip}
 \hline
 \noalign{\smallskip}
 18--200~keV & {\sc pulses} & pulse & -- & $39.0\pm0.8$ & $8.5\pm1.0$ & $15.0\pm1.7$ &  $0.29\pm0.05^{\mathrm{b}}$ & $1.08\pm0.13$\\
$H$ & {\sc pulses} & 1$^{\rm st}$ pulse & -- & $119.0\pm2.2$ & $18.2\pm4.8$ & $63.6\pm5.2$ & $4.9\pm0.3$ & $[1.08]$\\
$H$ & {\sc pulses} & 2$^{\rm nd}$ pulse & -- & $180.5\pm9.4$ & $0.5\pm3.6$ & $110\pm15$ & $2.6\pm0.2$ & $[1.08]$\\
$R$ & {\sc pulses} & 1$^{\rm st}$ pulse & -- & $[119.0]$  & $[18.2]$ & $[63.6]$ & $2.7\pm0.2$ & $[1.08]$\\
$R$ & {\sc pulses} & 2$^{\rm nd}$ pulse & -- & $[180.5]$ & $[0.5]$ & $[110]$ & $1.4\pm0.1$ & $[1.08]$\\
\\
$H$ & {\sc pl+pulses} & pl                 &$1.06\pm0.08$ & -- &  --         & --           & $3.6\pm2.0^{\mathrm{c}}$ & --\\ 
$H$ & {\sc pl+pulses} & 1$^{\rm st}$ pulse & -- & $119.0\pm0.1$ & $[0.1]$ & $27\pm18$ & $2.2\pm1.0$ & $[1.08]$\\
$H$ & {\sc pl+pulses} & 2$^{\rm nd}$ pulse & -- & $181.1\pm0.6$ & $[0.1]$ & $41\pm17$ & $2.6\pm0.8$ & $[1.08]$\\
$R$ & {\sc pl+pulses} & pl                 &$[1.06]$ & -- &  --         & --           & $2.0\pm1.1^{\mathrm{c}}$ & --      \\
$R$ & {\sc pl+pulses} & 1$^{\rm st}$ pulse & -- & $[119.0]$ & $[0.1]$ & $[27]$ & $1.2\pm0.6$ & $[1.08]$\\
$R$ & {\sc pl+pulses} & 2$^{\rm nd}$ pulse & -- & $[181.1]$ & $[0.1]$ & $[41]$ & $1.4\pm0.4$ & $[1.08]$\\
\\
late $R$ & {\sc pl+pulse} & pl                 &$[1.06]$ & -- &  --         & --           & $11.0\pm0.9^{\mathrm{d,e}}$ & --     \\ 
late $R$ & {\sc pl+pulse} & pulse & -- & $19.1\pm0.3^{\mathrm{f}}$  & $2.85\pm0.54^{\mathrm{f}}$ & $7.4\pm0.7^{\mathrm{f}}$ & $36\pm4^{\mathrm{e}}$ & $0.91\pm0.34$\\
0.3--10~keV & {\sc pl+pulse} & pl                 &$1.47\pm0.20$ & -- &  --         & --           & $1.3\pm0.1^{\mathrm{d,g}}$ & --     \\ 
0.3--10~keV & {\sc pl+pulse} & pulse & -- & $10\pm3^{\mathrm{f}}$  & $2.9\pm2.5^{\mathrm{f}}$ & $6.6\pm0.9^{\mathrm{f}}$ & $24\pm12^{\mathrm{g}}$ & $0.89\pm0.56$\\
 \noalign{\smallskip}
  \hline
  \end{tabular}
  \end{center}
  \begin{list}{}{} 
  \item[$^{\mathrm{a}}$] Time since GRB onset (corresponding to 45~s before the trigger time).
  \item[$^{\mathrm{b}}$] Flux density corresponding at 88~keV.
  \item[$^{\mathrm{c}}$] Flux at $t=100$~s.
  \item[$^{\mathrm{d}}$] Flux at $t=10^5$~s.
  \item[$^{\mathrm{e}}$] Units of $\mu$Jy.
  \item[$^{\mathrm{f}}$] Units of $10^4$~s.
  \item[$^{\mathrm{g}}$] Units of $10^{-13}$~erg~cm$^{-2}$~s$^{-1}$.
  \end{list}   
\end{table*}

%%%%%%%%%%%%%%%%%%%%%%%%%%%%%%%%%%%%%%%%%%%%%%%%%%%%%%%%%%%%%%%%%%%%%%%%%%%%%%%%%

%---------------------------------------------
\subsection{SED evolution}
\label{sec:SED}
\begin{figure}
\centering
\includegraphics[width=8.5cm]{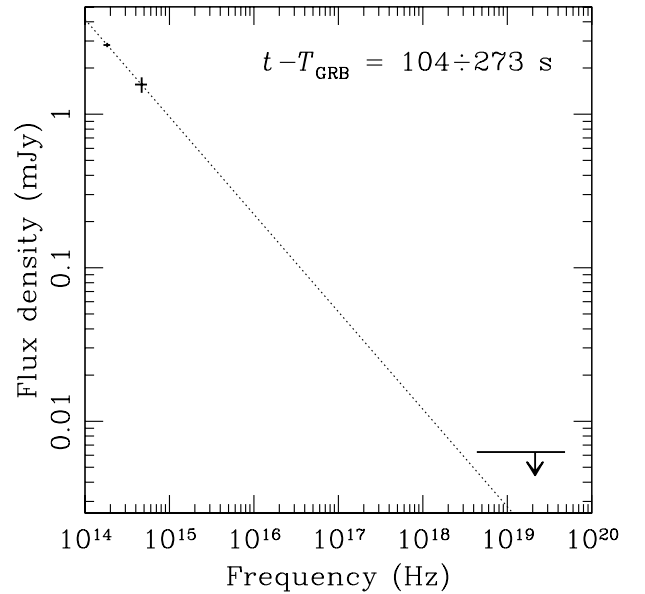}
\caption{NIR-optical/$\gamma$-ray SED between 104 and 273~s after the
GRB onset time. The dotted line shows the best-fit power law,
with $\beta_{\rm NIR-opt}=0.65\pm0.21$.
The upper limit is at 3~$\sigma$.
}
\label{fig:earlysed}
\end{figure}

%---------------------------------------------
We derived two SEDs at different epochs.
The early one comprises the NIR pulses seen at the beginning of the REM follow-up observations
and lasts from 104 to 273~s.
This SED consists of detections in two filters, $H$ and $R$, and a 3-$\sigma$ upper limit
of $6.3$~$\mu$Jy in the 18--200~keV band (see Fig.~\ref{fig:all_lc}).
Given the high variability of the $H$ curve, especially
when compared with that of $R$, the mean $H$ flux was derived by integrating the best-fit
profile of the $H$ curve described in Sect.~\ref{sec:pulses} over this time interval.
The simultaneous mean $R$ flux was calculated assuming the best-fit rescaling factor
($f_{\rm best}=0.55\pm0.06$) between $H$ and $R$ derived in Sect.~\ref{sec:multi_lc}
for the same time interval.
The fit with a simple power law yields $\beta_{\rm NIR-opt}=0.65\pm0.21$.
The extrapolation of the fit to the $\gamma$-ray band is consistent with the upper limit
(Fig.~\ref{fig:earlysed}).

Interestingly, the spectral index is consistent with that at the end of the $\gamma$-ray pulse
(Sect.~\ref{sec:gamma}), $\beta_\gamma=0.45\pm0.15$, and might be suggestive of an unbroken power
law segment from NIR to $\gamma$ at this epoch.
However, the large value of $E_{B-V}$ makes the uncertainty $\delta A_V$ on the
Galactic extinction quite large, $\delta A_V\ge0.5$~mag. Should the Galactic extinction
in $V$ be larger (smaller) by 0.5, the corrected spectral index of the NIR/optical data alone
would be $\beta_{\rm NIR-opt}=0.3\pm0.2$ ($\beta_{\rm NIR-opt}=1.0\pm0.2$).
Thus, nothing conclusive can be said about the possible presence of extinction in excess of the 
Galactic one.

\begin{figure}
\centering
\includegraphics[width=8.5cm]{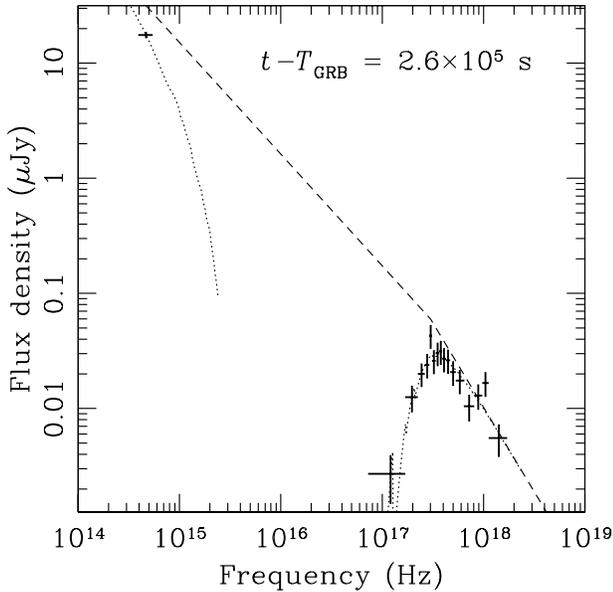}
\caption{Optical/X-ray SED at $2.6\times10^5$~s, around the peak
of the late rebrightening. The dashed (dotted) line shows the best-fit absorbed broken power law,
corrected (not corrected) for the optical extinction and X-ray absorption.
The $N_{\rm H}$ was fixed to $5.5\times10^{21}$~cm$^{-2}$ found from fitting the
X-ray spectrum alone, while $A_V$ was found to be $0.80\pm0.15$~mag, adopting an SMC profile
(at $z=0$). 
}
\label{fig:latesed}
\end{figure}

The late SED (Fig.~\ref{fig:latesed}) was extracted at $2.6\times10^5$~s, corresponding to the
beginning of the decay following the late rebrightening in $R$ simultaneously with the final steep
X-ray decay (see Fig.~\ref{fig:all_lc}). This SED includes a single $R$ measurement taken by
\citet{Halpern07d} and the XRT spectrum BCD (Sect.~\ref{sec:X_spec}; Table~\ref{tab:xspec})
renormalised through the best-fit power-law segment of the X-ray light curve (Sect.~\ref{sec:X_lc})
at the epoch of the $R$ point.
The SED was fit with an SMC-extinguished \nocite{Pei92}(Pei 1992; in the observer frame),
X-ray photoelectrically absorbed broken power law
with $\beta_{\rm x}-\beta_{\rm NIR-opt}=0.5$. The $N_{\rm H}$ was fixed to the value already
derived from the corresponding X-ray spectrum, i.e. $5.5\times10^{21}$~cm$^{-2}$,
consistent with being entirely Galactic (Sect.~\ref{sec:X_spec}).
We found $\beta_{\rm NIR-opt}=1.0\pm0.2$, consistent with $\beta_{\rm x}=1.5$
from the X-ray spectrum best fit (Table~\ref{tab:xspec}). The break frequency turned out to be
$\nu_{\rm b}=(3.0\pm0.9)\times10^{17}$~Hz and $A_V=0.80\pm0.15$~mag ($\chi^2/{\rm dof}=9.6/12$).
Figure~\ref{fig:latesed} shows the SED: the dashed (dotted) line represents the best-fit model
with optical extinction and X-ray absorption taken out (shown).
%
% ++++++++++++++ Early SED ++++++++++++++
%++++++++++++++++++++++++++++++++++++++++++
%
%
% ++++++++++++++ Late SED ++++++++++++++
%++++++++++++++++++++++++++++++++++++++++++
%

Alternatively, we tried to fit the SED with a single absorbed power law.
If we left the slope free to vary, the fit is driven by the more numerous
X-ray points, leading to $\beta\sim1.5$ and  $A_V=6\pm3$~mag.
However, if we correct the early SED for such an extinction, the intrinsic
optical spectrum would be nonphysically blue ($\beta_{\rm opt}=-3$).
Otherwise if we impose a single power law between optical and X-ray and
fix $\beta_{\rm opt-x}=1.0$, $A_V$ becomes comparable
with that found in the case of a broken power law. However,
although the fit cannot be rejected ($\chi^2/{\rm dof}=18/13$),
the residuals of the X-ray points with respect to the model show a trend.
Therefore, a simple absorbed power law does not seem to be a good representation
of our data.

Because of the unknown redshift $z$ of GRB~070311, we caution that the values of $A_V$ computed
in the fits above are calculated for $z=0$ assuming a specific extinction law (SMC model),
so they must be taken as indicative upper limits to the corresponding rest-frame values. 
Unlike for the early SED, the X-ray absorption in terms of $N_{\rm H}$ is known to be consistent
with the expected Galactic value.
Due to the unknown redshift, it is not possible to set an upper limit to
the possible intrinsic rest-frame $N_{{\rm H},z}$ absorption.
For this reason and because of the upper limits on $A_V$, nothing can be inferred on the
amount of dust and gas along the line of sight to the GRB progenitor.

%%%%%%%%%%%%%%%%%%%%%%%%%%%%%%%%%%%%%
\section{Discussion}
\label{sec:disc}
%%%%%%%%%%%%%%%%%%%%%%%%%%%%%%%%%%%%%
Together with the early optical/NIR flares possibly accompanied by a $\gamma$-ray tail,
the late, bright and long rebrightening seen in X-ray and $R$ band probably is the most intriguing
feature of GRB~070311.
Although such late brightenings or flares are not unprecedented, only a few bursts have exhibited them
so far: e.g., GRB~970508 \citep{Piro98,Galama98},
the short GRB~050724 \citep{Campana06,Malesani07}, the $z=6.3$ GRB~050904 \citep{Cusumano07,Watson06}.
We also note the presence of the shallow decay phase preceding the X-ray brightening (or corresponding to
its gradual onset). Due to the lack of contemporaneous coverage in the $R$ band, we did not observe a
similar shallow decay phase in the optical bands. However, from Fig.~\ref{fig:all_lc} 
we infer that it must have taken place (e.g., some energy injection between $\sim10^4$~s and $\sim10^5$~s)
to power the continuum component of the $R$-filter decay at late times, which is
significantly above the extrapolation of the early data.

In the case of GRB~070311, we note that $\Delta t/t\sim1$ for the late rebrightening in both X-ray and
$R$ filter, so timescales arguments cannot be used against an external shock origin for it.
We note that this is in contrast to what has been observed for other GRBs
(e.g. GRB~050502B; Burrows et al. 2005,\nocite{Burrows05} Falcone et al. 2006;\nocite{Falcone06}
GRB~050724; Barthelmy et al. 2005,\nocite{Barthelmy05} Campana et al. 2006),\nocite{Campana06}
in which the rise and decay of the flares are too steep and require a resetting of the time origin.

In addition, the FRED-like shape of the 18--200~keV prompt light curve and consequent lack of high
variability, classically used to argue against an external origin of the prompt emission
of complex bursts (Sari \& Piran 1997; but see also Dermer \& Mitman 1999),\nocite{Sari97,Dermer99}
indicates that the prompt emission of GRB~070311 is potentially consistent with it
(e.g., see Kumar \& Panaitescu 2003),\nocite{Kumar03} although
unlikely due to synchrotron radiation \citep{Ramirez06}.

Motivated by this and by the analogies found between prompt and late afterglow, 
hereafter we try to interpret the observations of GRB~070311 in the light of an external
shock origin. While the $\gamma$-ray pulse is thought
to be produced during the deceleration of the shell against the surrounding medium, we consider
two alternative explanations for the late rebrightening: either a refreshed shock by a second
shell emitted after the first one, or a single shell whose forward shock encounters
a density bump.

The description of the $R$ and X-ray late rebrightening as a simple power law with a slope
changing after the peak (Sect.~\ref{sec:late_pl_pulse}), is consistent with the interpretation
that the late rebrightening is due to a thin shell that caught up with the shock front of
the blastwave at later times, as expected in the so-called refreshed-shock scenario \citep{Rees98}.
In this picture, the fireball rebrightened and soon afterwards ($\sim 2\times10^5$~s) turned off:
the decay is the result of two components: the pre-existing power law and the high-latitude
radiation left over by the refreshed shock.
This is the case when the energy of the impacting shell is lower than that of the fireball
(as also suggested by the ratio between the radiated energy during the late flare and the prompt emission),
so that no noticeable step in the power-law decay is observed, but still luminous enough
to produce a peak on a short timescale.

According to the so-called ``curvature'' effect \citep{Fenimore96,Kumar00,Dermer04}, the steepness
of the fast decay phase of several X-ray afterglows as well as of the $\gamma$-ray tail of single
pulses is due to high-latitude radiation Doppler-boosted in the observer fixed energy band.
A relation is expected between temporal and spectral index: $\alpha=\beta+2$ in its simplest form
in the case of a thin shell emitting for a short time, where the time origin must correspond to
the time of the GRB explosion as measured in the detector frame.

Notably, for GRB~070311 both the decay of the $\gamma$-ray pulse and the longer
decays in X-rays and $R$-filter of the late rebrightening are consistent with the high-latitude
closure relation and {\em with the same time origin},
i.e. the GRB onset time: $\alpha_\gamma\sim\beta_\gamma+2=2.3\pm0.1$
and $\alpha_{\rm x}=\beta_{\rm x}+2=3.5\pm0.2$.
This is proven by the measured slope of
the second power law, $\alpha_{\rm x,2}^{\rm (pp)}=3.5\pm0.5$, which describes the flare decay, in addition
to the underlying power law with $\alpha_{\rm x,1}^{\rm (pp)}=1.4\pm0.1$ (Sect.~\ref{sec:late_pl_pulse}).
Therefore, if we adopt the combination of a power law and either a pulse or another power law
for the late rebrightening (Sect.~\ref{sec:late_pl_pulse}), the slope of the pulse decay is
still consistent with a curvature effect: $\alpha_{\rm x,2}^{\rm (pp)}=\beta_{\rm x} + 2$
In addition, from Sect.~\ref{sec:gamma_late_pl_pulse} we know that if we move the reference time
backward to $19.0\pm2.4$~s before the GRB onset, the late brightening is well fit by a time-rescaled version
of the prompt $\gamma$-ray pulse. What is remarkable is that the peak time scales exactly in the same
way as the pulse rise and decay times:
$\Delta t_{\gamma,{\rm prompt}}/t_{\gamma,{\rm prompt}}=\Delta t_{\rm x,late}/t_{\rm x,late}$.
This suggests that either two shells generated the prompt pulse as well as the late rebrightening or,
alternatively, that a unique shell caused both by encountering two density enhancements.
We note that the late X-ray hump occurs later (few $\times10^4$~s) than what is more commonly seen
in the typical flat phase ($\sim10^3$~s), also usually interpreted in terms of energy injection
from the central engine. The detection in the $R$ filter rules out a very high redshift,
so such a late refreshing may be either due to a more distant radius of emission of the forward
shock (e.g. due to low density of the surrounding medium) or due to a slow shell catching up with the
shock front.

The 18--200~keV fluence emitted in the prompt amounts to $(3.0\pm0.5)\times10^{-6}$~erg~cm$^{-2}$
by integrating the best-fit model of Sect.~\ref{sec:pulses},
while the net 0.3--10~keV fluence of the late rebrightening is $(2.4\pm1.2)\times10^{-7}$~erg~cm$^{-2}$,
i.e. lower by one order of magnitude. If we extrapolate the prompt spectrum to the XRT band
as in Sect.~\ref{sec:pl_pulses} and correct for the X-ray absorption, the fluence of the late X-ray
rebrightening becomes comparable with that of the prompt extrapolated to the same energy band.
Differently, the late $R$ hump has a time-integrated flux about one order of magnitude larger
than the early pulses seen in the same filter.
The larger energy content of the late shock, with respect to the early one seen in optical,
might be explained with an increasing efficiency in converting the kinetic energy of the
blastwave into radiation. This implies a change of the microphysical parameters of the afterglow, as
suggested to explain the shallow decay phase of other GRBs \citep{Ioka06,Panaitescu06}.
This can also be explained more simply as due to the fact that the $\nu\,F_\nu$ broadband spectrum
peaks at lower energies at later times, so that the energy release in the observed $R$ filter
during the late rebrightening is larger than during the prompt or soon after that.

Following \citet{Ioka05}, we can find which scenarios may be compatible with the observed
flux increase, $\Delta F/F\approx 10$ (both in $R$ and X; Sect.~\ref{sec:late_pl_pulse}),
and the observed $\Delta\,t/t\approx1$: from their fig.~1, the late hump of GRB~070311 is
compatible with the refreshed shock scenario. Also the patchy shell model
\citep{Meszaros98}, characterised by an anisotropic emitting surface of the fireball,
is not ruled out, while the scenario of a density bump causing the late rebrightening
is ruled out, unless many clumps of matter are illuminated at the same time.
Therefore, the bright flux of the rebrightening, $\Delta F/F\approx 10$, seems to disfavour
the interpretation of a density medium enhancement, causing
the observed flux increase. However we note that this is debated: e.g., according to
Dermer \& Mitman (1999) and Dermer (2007a),\nocite{Dermer99,Dermer07a}
this could be produced by parts of the blastwave
in which most of the energy has not been converted into radiation, yet, while impacting on a dense and
thick clump of matter. In this scenario, the same shell would be responsible for both
the prompt and the late rebrightening: e.g., the high-latitude emission observed during the decay
of the late hump might be the result of the blastwave finally reaching the wind-termination
shock.
However, the interpretation of the late rebrightening as due to a density bump seems disfavoured
by the crossing of the X-ray band by the cooling break $\nu_{\rm c}$,
whereas the observed frequency must be below $\nu_{\rm c}$.
In addition to that, the remarkable flux enhancement observed, $\Delta\,F_{\rm x}/F_{\rm x}=18\pm9$
(Sect.~\ref{sec:late_pl_pulse}), makes this interpretation less probable.
In both scenarios (refreshed shock produced by another shell or density bump),
the scaling factor of the timescale of the late brightening with respect to the early pulse,
$f_{\rm s}=5700\pm700$, could result from the interplay of two factors:
the increase of the visible portion of the emitting surface and the fact that the blastwave
Lorentz factor has decreased by a factor of $\Gamma_1/\Gamma_2$ head on, thus stretching the
timescale by $(\Gamma_1/\Gamma_2)^2$.
Therefore, since $R_1<R_2$, from $f_{\rm s}=R_2/R_1\,(\Gamma_1/\Gamma_2)^2$, it must be
$\Gamma_1/\Gamma_2<\sqrt{f_{\rm s}}=75\pm5$.

The early NIR/optical pulses detected soon after the $\gamma$-ray pulse are consistent
with both interpretations: they could be the result of density bumps swept up by the
blastwave ($\nu_{\rm NIR/opt}<\nu_{\rm c}$), or other shells emitted soon after the first
one and catching up with the shock front 80 and 140~s after its deceleration
(Sect.~\ref{sec:dec_radius}).

During the late rebrightening, we know that the spectrum is likely described by a broken
power law with $\beta_{\rm x}=1.5\pm0.2$ and $\beta_{\rm opt}=1.0\pm0.2$.
If we interpret it as the cooling break in the slow cooling regime, the electron power-law
distribution index is $p=3.0\pm0.4$. The fact that $\beta_{\rm x}$ was around 1.0 at the
beginning of the XRT observations can be explained if the cooling frequency crossed the X-ray
band during the observations. The change in the X-ray decay that would be implied amounts
to 1/4 and could be still compatible with the pre-break slope, ranging from 1.0 to 1.4, depending
on which model one assumes (Sects.~\ref{sec:X_lc}, \ref{sec:pl_pulses}, \ref{sec:late_pl_pulse}).
The optical decay $\alpha_R=1.06\pm0.08$ is compatible with that expected in the case of
an ISM environment: $\alpha_R^{\rm ISM}=3(p-1)/4=1.5\pm0.3$. Differently, the case of a
wind environment is ruled out (3~$\sigma$): $\alpha_R^{\rm wind}=(3p-1)/4=2.0\pm0.3$.

Alternatively, if during the late rebrightening the bulk Lorentz factor has already decreased
to $\Gamma<1/\theta_{\rm j}$, where $\theta_{\rm j}$ is the jet opening angle, the afterglow
should already have experienced an achromatic jet break: in this case, both optical and X-ray
decay indices are simply equal to $p$. Notably, this is compatible with the measured values
during the decay of the late rebrightening and this led \citet{Panaitescu07} to favour the
jet interpretation for this burst.
However, we believe that the late rebrightening is more likely to be due to an energisation
of the blastwave shock front that strongly affected the measured power-law slope,
similarly to what was inferred in the case of GRB~050724 \citep{Malesani07}.

The interpretation of the final steep decay following the flare as the post jet-break
decay seems unlikely: while the late X-ray flare might be still interpreted
as a shallow-steep transition (Sect.~\ref{sec:X_lc}), in the optical the identification
as a flare cannot be questioned. Therefore, the steep decay after the peak simply
corresponds to the declining part of the flare and not to a jet break.

%--------------------------------
\subsection{Deceleration radius of the fireball}
\label{sec:dec_radius}
%--------------------------------
The result of Sect.~\ref{sec:gamma_late_pl_pulse} suggests that the time origin, moved to
$19.0\pm2.4$~s before the GRB onset, would correspond to the explosion time in the detector
frame (hereafter DF), i.e. when the shell radii are negligible.
Let $t_{\rm expl}=-19.0$~s be the explosion time measured in the DF.
In this context we can derive some clues on the fireball evolution produced during the first
shock corresponding to the $\gamma$-ray pulse.
In this picture the shell would expand from $t_{\rm expl}$ to $t_{\gamma}=0$, when it would
start emitting $\gamma$ rays. The deceleration time $t_{\rm dec}$ would correspond to
the peak time of the $\gamma$-ray pulse, i.e. $39.0\pm0.8$~s.
First the shell accelerates until it reaches the coasting radius (typically $\sim10^{13}$~cm),
while the bulk Lorentz factor increases linearly with radius until it reaches the maximum value of
$\Gamma_0$. At this stage, the internal energy of the fireball has been converted into bulk
kinetic energy. After that, the shell expands with constant $\Gamma=\Gamma_0$ until it is
decelerated by the surrounding medium.
If the time it takes the shell to reach the coasting phase is negligible with
respect to the time it takes to begin to decelerate, it follows that during most of the time
from $t_{\rm expl}$ to $t_{\gamma}$ the shell was moving with $\Gamma_0$.
In this case the shell would start decelerating at the radius
$2\,c\,(t_{\gamma}-t_{\rm expl})\,\Gamma_0^2/(1+z)=1.1\times10^{16}\,(\Gamma_0/100)^2\,(1+z)^{-1}$~cm.
The deceleration process would culminate at the deceleration radius
$r_{\rm dec}=(3\,E_{\rm iso}/4\pi\,n\,m_{\rm p}c^2\Gamma_0^2)^{1/3}\simeq (5.4\times10^{16}~{\rm cm})\,(E_{\rm iso,52}/n)^{1/3}\,(\Gamma_0/100)^{-2/3}$ at the deceleration time
(DF) $t_{\rm dec}=94\,(E_{\rm iso,52}/n)^{1/3}\,(\Gamma_0/100)^{-8/3}\,(1+z)$~s in the
thin shell case ploughing into uniform ISM \citep{Rees92}.
We note that from the beginning to the peak of the deceleration, i.e. from $t_\gamma$ to
$t_{\rm dec}$, the bulk Lorentz factor decreases approximately from $\Gamma_0$ to $\Gamma_0/2$.
From this we can infer the distance travelled by the fireball in the same time interval:
this must be between $2\,c\,(t_{\rm dec}-t_{\gamma})\,(\Gamma_0/2)^2/(1+z)$
and $2\,c\,(t_{\rm dec}-t_{\gamma})\,\Gamma_0^2/(1+z)$, i.e. between
$0.6\times10^{16}\,(\Gamma_0/100)^2\,(1+z)^{-1}$~cm and 
$2.3\times10^{16}\,(\Gamma_0/100)^2\,(1+z)^{-1}$~cm. This is comparable with the distance already
travelled at the beginning of the deceleration and consistent with the numbers reported above.
The unknown redshift $z$ unfortunately makes it too tentative to push these estimates any further.

%%%%%%%%%%%%%%%%%%%%%%%%%%%%%%%%%%%%%
\section{Conclusions}
\label{sec:conc}
%%%%%%%%%%%%%%%%%%%%%%%%%%%%%%%%%%%%%
GRB~070311 is a FRED-shaped burst followed by early NIR/optical pulses detected from
$\sim100$ to $\sim200$~s and possibly accompanied by a simultaneous faint $\gamma$-ray tail,
 with subsequent variability detected at different NIR/optical filters up to $\sim10^3$~s since the GRB onset.
Another remarkable property exhibited by GRB~070311 is the late $R$ and X-ray rebrightening
observed around few $10^5$~s after the burst, with the X-ray peaking earlier than $R$-filter photons.
When we fit it with the combination of an underlying power law plus a pulse, 
the X-ray ($R$) fluence of the pulse alone is comparable (10 times larger) with that of the early pulse,
while the peak intensity is about one order of magnitude larger than that of the underlying power law.
Interestingly, if we refer the times to $19.0\pm2.4$~s prior to the GRB onset, it turns out
that the peak time as well as the rise and decay times of the late pulse are compatible with
the corresponding times of the $\gamma$-ray pulse, rescaled by the same factor: $f_{\rm s}=5700\pm700$.
We interpreted this in the external shock scenario, where the $\gamma$-ray prompt emission
would correspond to the deceleration of the blastwave sweeping up the surrounding medium
with uniform density, while the late $R$ and X-ray rebrightening would be produced by the
refreshed shock of another shell emitted after the first and impacting the blastwave when this
has a Lorentz factor $\Gamma_2$, so that:  $\Gamma_1/75<\Gamma_2<\Gamma_1$.
In this context, the time offset of $19.0\pm2.4$~s before the GRB would correspond to the explosion
time in the detector rest frame, while the GRB onset would mark the beginning of the $\gamma$-ray
emission due to the deceleration of the fireball. From the explosion, onset and peak times
we infer consistent estimates of the deceleration radius,
a few $\times10^{16}\,(\Gamma_0/100)^2\,(1+z)^{-1}$~cm.

The interpretation of the late rebrightening as the result of a density bump in the surrounding
medium would explain naturally both the prompt and the late hump with a single shell.
However, the possible presence of the cooling break close to the X-ray band around the peak time,
combined with the remarkable flux enhancement observed, $\Delta\,F/F\approx10$, makes this
scenario less appealing.

The occurrence of the early NIR/optical flares at 80 and 140~s after the peak of the $\gamma$-ray
emission is consistent with both scenarios: either due to density enhancements of the
matter encountered by the blastwave or explained by further shells catching up with it.

Finally, we note that the early (late) NIR/optical pulses do not peak contemporaneously with
the corresponding $\gamma$-ray (X-ray) pulses, but are delayed by a factor of $\sim3$ ($\sim2$)
in time. A clear interpretation of this effect is lacking.

%%%%%%%%%%%%%%%%%%%%%%%%%%%%%%%%%%%%%

\acknowledgements{}

This work is supported by ASI grant I/R/039/04 and
by the Ministry of University and Research of Italy
(PRIN 2005025417). S.D.V. is supported by SFI grant 05/RFP/PHY0041.
D.M. acknowledges the Instrument Centre for
Danish Astrophysics for financial support.
The Dark Cosmology Centre is funded by the DNRF.
We gratefully acknowledge the contribution of dozens 
of members of the XRT team at OAB, PSU, UL, GSFC,
ASDC and our sub-contractors, who helped make 
this instrument possible.
INTEGRAL is an ESA project funded by ESA member states
(especially the PI countries: Denmark, France, Germany, Italy,
Switzerland), Czech Republic and Poland, and with the participation
of Russia and the USA.
This study is supported by Spanish research programmes
ESP2002-04124-C03-01 and AYA2004-01515. Partially based on observations
collected at the German-Spanish Astronomical Center, Calar Alto, jointly
operated by the Max-Planck-Institut f\"ur Astronomie Heidelberg and the 
Instituto de Astrof\'isica de Andaluc\'ia (CSIC).
%%%%%%%%%%%%%%%%%%%%%%%%%%%%%%%%%%%%%%

\clearpage

\end{document}